\documentclass[a4paper,reqno,12pt,final]{amsart}
\usepackage{amsmath,amssymb,eucal,bbold}
\usepackage{subfigure}
\usepackage[latin1]{inputenc}
\usepackage{array}
\usepackage{epic,eepic}
\usepackage{color}
\definecolor{light}{gray}{0.85}
\definecolor{light2}{gray}{0.60}
\newcommand{\goodgap}{%
  \hspace{\subfigtopskip}%
  \hspace{\subfigbottomskip}}
\calclayout
\renewcommand{\d}{\partial}
\newcommand{\into}{\hookrightarrow}
\newcommand{\half}{\tfrac12}
\newcommand\la{\langle} 
\newcommand\ra{\rangle}

\newcommand{\codim}{\text{codim}}

\renewcommand{\gg}{\mathfrak g}
\newcommand{\gh}{\mathfrak h}
\newcommand{\gt}{\mathfrak t}
\newcommand{\1}{\mathbb 1}

\newcommand{\Z}{\mathbb Z}

\newcommand{\su}{\mathfrak{su}}
\newcommand{\so}{\mathfrak{so}}

\newcommand{\eC}{\mathcal{C}}

\newcommand{\eG}{\mathcal{G}}

\newcommand{\eM}{\mathcal{M}}
\newcommand{\eO}{\mathcal{O}}

\newcommand{\eZ}{\mathcal{Z}}

\newcommand{\Aut}{\mathrm{Aut}}
\newcommand{\Inn}{\mathrm{Inn}}
\newcommand{\Out}{\mathrm{Out}}

\newcommand{\U}{\mathrm{U}}
\newcommand{\SU}{\mathrm{SU}}
\newcommand{\SO}{\mathrm{SO}}
\renewcommand{\O}{\mathrm{O}}
\newcommand{\ZZ}{\mathbb{Z}}
\DeclareMathOperator{\rank}{rank}

\DeclareMathOperator{\Ad}{Ad}
\DeclareMathOperator{\Tr}{Tr}

\begin{document}

\title[]{An illustrated guide to D-branes in $\SU_3$}
\author[]{Sonia Stanciu}
\address[]{\begin{flushright}Spinoza Institute\\
Utrecht University\\
Leuvenlaan 4\\
 Utrecht\\
The Netherlands\\
\end{flushright}}
\address[]{\begin{flushright}Theoretical Physics Group,\\
Blackett Laboratory,\\ Imperial College,\\
Prince Consort Road,\\ London SW7 2BZ\\
U.K.\\
\end{flushright}}
\email{s.stanciu@phys.uu.nl, s.stanciu@ic.ac.uk} 
\thanks{${}^*$ SPIN-2001/22, Imperial/TP/01-2/2, {\tt hep-th/0111221}} 
\date{\today}
\begin{abstract}
  We give a systematic account of symmetric D-branes in the Lie group
  $\SU_3$.  We determine both the classical and quantum moduli space
  of (twisted) conjugacy classes in terms of the (twisted) Stiefel
  diagram of the Lie group.  We show that the allowed (twisted)
  conjugacy classes are in one-to-one correspondence with integrable
  highest weight representations of the (twisted) affine Lie algebra.
  In particular, we show how the charges of these D-branes fit in the
  twisted K-theory groups.
\end{abstract}
\maketitle

\section{Introduction}

Group manifolds provide an ideal laboratory for the study of D-branes
in general backgrounds, as they are amenable to a variety of
approaches, ranging from the algebraic techniques of BCFT to the
lagrangian description based on the boundary WZW model.  At present,
the simplest and best understood class of D-brane configurations is
obtained \cite{AS,FFFS,SDnotes} as solutions of the familiar gluing
conditions on the chiral currents of the WZW background
\begin{equation}\label{eq:gc}
J(z) = R \bar J(\bar z)~,
\end{equation}
where $R$ is a metric preserving Lie algebra automorphism.  They
describe \emph{symmetric} D-branes, by which we mean D-branes wrapping
(twisted) conjugacy classes in the group manifold $G$.  Both standard
and twisted conjugacy classes are homogeneous spaces in the group
manifold.  In fact, it is often stated that `most' D-brane
configurations wrap conjugacy classes of `regular' elements.  This
assertion is based on the fact that in a group manifold, regular
points are dense while singular points form a set of codimension at
least three.  This seems enough to justify restricting ourselves to
the above (twisted) conjugacy classes, as the corresponding D-brane
configurations are indeed generic at the classical level.  At the
quantum level, however, one has a finite number of D-brane
configurations which always include conjugacy classes of `singular'
elements.  It therefore appears desirable to have a complete picture
of all the possible D-brane configurations.  The main aim of this
paper is to give a such a complete picture of the consistent D-brane
configurations on (twisted) conjugacy classes in the particular case
of the $\SU_3$ group manifold, leaving the general analysis for a
forthcoming paper \cite{SD2notes}.  D-branes in the $\SU_3$ manifold
have been analysed recently in \cite{FredSch,MMS2}, using various
approaches.  Here we would like to offer a complementary description,
from a geometric point of view, paying special attention to the
twisted case.

The paper is organised as follows.  Section~2 is devoted to the case
$R=\1$.  In this case, we provide a natural description of the space
of D-branes in terms of the Stiefel diagram of $\SU_3$.  This is a
figure in the Cartan subalgebra describing the singular points of the
group in a maximal torus.  The moduli space of classical D-branes is
described by the fundamental domain of the extended Weyl group which
is given by an equilateral triangle with the interior points
describing `regular' six-dimensional D-branes, whereas the boundary
points describe lower-dimensional D-branes.  The requirement of
single-valuedness of the path integral of the boundary WZW model
selects a finite number of consistent configurations at every given
level, each of them being uniquely characterised by its intersections
with certain cosets of the $\SU_2^\alpha$ subgroups corresponding to
the simple roots of $\SU_3$.

In preparation for our study of the quantisation conditions for
twisted conjugacy classes, in Section~3 we briefly describe the
quantisation conditions in the case of the $\SO_3$ WZW model, which
will prove instrumental in our analysis of the twisted case for
$\SU_3$.  We compare the space of D-submanifolds of the $\SO_3$ and
$\SU_2$ theories, both at the classical and quantum levels.

In Section 4 we undertake the study of the D-brane configurations
wrapping twisted conjugacy classes.  Describing these conjugacy
classes as orbits in a non-connected extension of $\SU_3$ allows us to
use the theory of non-connected Lie groups in order to describe their
moduli space.  To this end we introduce a generalisation of the
Stiefel diagram and the notion of \emph{twisted} regularity, which is
more appropriate for the description of twisted conjugacy classes.  We
analyse the quantisation conditions imposed by the condition of
single-valuedness of the path integral and we obtain a discrete
spectrum of consistent D-brane configurations.  We show that the
admissible D-branes are in one-to-one correspondence with the
integrable highest weight (IHW) representations of the twisted
affine Lie algebra $\widehat\su(3)^{(2)}_k$.  We describe the way in
which the admissible twisted branes in $\SU_3$ intersect the fixed
point subgroup $\SO_3$ in conjugacy classes which describe admissible
D-branes of a restricted WZW model.

In Section~5 we discuss the charges of the D-branes in $\SU_3$ and
their relation to the twisted K-theory group $K_H(\SU_3)$ which was
analysed in \cite{MMS2}.  For the untwisted branes, our results agree
with the ones obtained previously in \cite{FredSch,MMS2}.  For the
twisted branes, we argue that the charge is given by the dimension of
the inducing representation of the fixed point subalgebra.  An explicit
calculation shows that the charges of the configurations determined in
Section 4 fit in the twisted K-theory group $K^1_H(\SU_3)$.  Our
approach offers support for the idea that the charges carried by
symmetric D-branes can be determined by entirely geometric and
topological means, without any dynamical input.

In Section~6 we discuss our $\SU_3$ results and we outline the general
picture that emerges from this analysis.  Finally, we include two
appendices.  In Appendix~\ref{sec:su32k} we collect the basic facts
about the twisted affine Lie algebra $\widehat\su(3)^{(2)}_k$ which we
use in our analysis of twisted branes.  In
Appendix~\ref{sec:homology}, written jointly with José
Figueroa-O'Farrill, we prove several results on the relative
(co)homology groups of $\SU_3$ modulo a twisted conjugacy class.  In
particular we determine which twisted conjugacy classes are
quantum-mechanically consistent, a result which is used in Section~4.

\section{Untwisted D-branes in $\SU_3$}

\subsection{Classical analysis}

We start with the simpler case where the gluing automorphism $R$ is
taken to be the identity.  In this case the classical moduli space
$\eM_{cl}:=\eM_{cl}(\SU_3,\1)$ of D-branes is the space of conjugacy
classes of $G = \SU_3$.  We recall that a conjugacy class $\eC(g)$ of
an element $g$ of $G$ is defined as the orbit of $g$ in $G$ under the
adjoint action $\Ad_k:g\mapsto kgk^{-1}$, for any $k$ in $G$.  Since
the stabiliser of $g$ is given, in this case, by its centraliser
$\eZ(g)$, the conjugacy class $\eC(g)$ can be described as the
homogeneous space
\begin{equation}\label{eq:cc}
  \eC(g)\cong G/{\eZ(g)}~.
\end{equation}

We know that any point $g$ in the group manifold is conjugate, via the
adjoint action, to a point $h$ in a (fixed) maximal torus $T$, which
is determined up to a Weyl transformation in $W$.  Furthermore every
such element $h$ is the image, under the exponential map, of an
element $X$ in the Cartan subalgebra $\gt$ of $\su(3)$, which is only
determined up to a translation in the integer lattice $\Lambda_I$.
(Recall that the integer lattice consists of the kernel of the
exponential map.)  Thus the space of conjugacy classes in $\SU_3$ is
described by the quotient
\begin{equation*}
  \eM_{cl} = \frac{\gt}{W\ltimes\Lambda_I}~.
\end{equation*}
Moreover, since $\SU_3$ is simply connected, its integral lattice
$\Lambda_I$ agrees with its coroot lattice $\Lambda^\vee_R$, and thus
the semidirect product in the denominator is nothing but the extended
Weyl group $\widetilde W = W\ltimes\Lambda^\vee_R$.

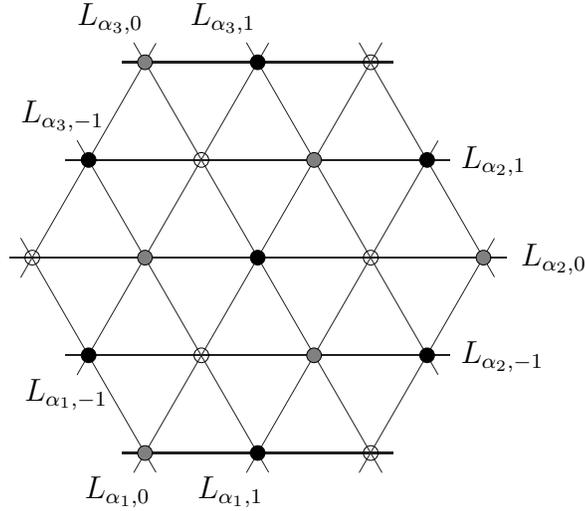
\begin{figure}[h!]
  \begin{center}
    \begin{picture}(80,80)
      \put(40,40){%
        \thinlines
        \put(-33,0){\line(1,0){66}}
        \put(-25.5, 12.9904){\line(1,0){51}}
        \put(-25.5, -12.9904){\line(1,0){51}}
        \put(-18, 25.9808){\line(1,0){36}}
        \put(-18, -25.9808){\line(1,0){36}}
        \put(-31.5, -2.59808){\line(56,97){18}}
        \put(-24, -15.5885){\line(56,97){25.5}}
        \put(-16.5, -28.5788){\line(56,97){33.}}
        \put(-1.5, -28.5788){\line(56,97){25.5}}
        \put(13.5, -28.5788){\line(56,97){18.}}
        \put(-31.5, 2.59808){\line(56,-97){18.}}
        \put(-24, 15.5885){\line(56,-97){25.5}}
        \put(-16.5, 28.5788){\line(56,-97){33.}}
        \put(-1.5, 28.5788){\line(56,-97){25.5}}
        \put(13.5, 28.5788){\line(56,-97){18.}}
        \put(0,0){\circle*{2}}
        \put(22.5, 12.9904){\circle*{2}}
        \put(0, 25.9808){\circle*{2}}
        \put(-22.5, 12.9904){\circle*{2}}
        \put(-22.5, -12.9904){\circle*{2}}
        \put(0, -25.9808){\circle*{2}}
        \put(22.5, -12.9904){\circle*{2}}
        \put(7.5, 12.9904){\shade\circle{2}}
        \put(-15., 0){\shade\circle{2}}
        \put(7.5, -12.9904){\shade\circle{2}}
        \put(30, 0){\shade\circle{2}}
        \put(-15, -25.9808){\shade\circle{2}}
        \put(-15, 25.9808){\shade\circle{2}}
        \put(15,25.9808){\circle{2}}
        \put(15,-25.9808){\circle{2}}
        \put(-7.5,12.9904){\circle{2}}
        \put(-7.5,-12.9904){\circle{2}}
        \put(15,0){\circle{2}}
        \put(-30,0){\circle{2}}
        }
      \put(-10,-10){%
        \put(85,49){$L_{\alpha_2,0}$}
        \put(77,62){$L_{\alpha_2,1}$}
        \put(77,36){$L_{\alpha_2,-1}$}
        \put(27,18){$L_{\alpha_1,0}$}
        \put(42,18){$L_{\alpha_1,1}$}
        \put(19,31){$L_{\alpha_1,-1}$}
        \put(26,81){$L_{\alpha_3,0}$}
        \put(41,81){$L_{\alpha_3,1}$}
        \put(18.5,68){$L_{\alpha_3,-1}$}
        }
    \end{picture}
    \caption{Stiefel diagram of $\SU_3$. The vertices correspond to
      the central lattice of $\SU_3$.}
    \label{fig:A2}
  \end{center}
\end{figure}

One can give a beautiful pictorial description of the space of
conjugacy classes in $\SU_3$ by using the Stiefel diagram (see, e.g.,
\cite{Adams}), which is represented in Figure~\ref{fig:A2}.  This is a
figure in the Cartan subalgebra $\gt$ describing the inverse image,
under the exponential map, of the singular points of a Lie group in
its maximal torus.  In our case, this consists of $3$ families of
$1$-dimensional hyperplanes: every positive root $\alpha_i$ in
$\Phi_+=\{\alpha_1,\alpha_2,\alpha_3=\alpha_1+\alpha_2\}$ gives rise
to a family $\{L_{\alpha_i, n}\}_{n\in\Z}$ of affine lines in $\gt$,
where each line $L_{\alpha_i, n}$ consists of the points $X$ in $\gt$
that satisfy
\begin{equation*}
L_{\alpha_i, n}~:\qquad\quad \alpha_i(X)=n~,\qquad\qquad i=1,2,3.
\end{equation*}
The set of regular points in $\gt$ decomposes in this way into convex
connected components, also known as the alcoves of the group.  The
group of isometries of $\gt$ generated by the reflections on the lines
$L_{\alpha_i, n}$ is nothing but the extended Weyl group of $\SU_3$.
We therefore see that the space of conjugacy classes in $\SU_3$ can
be identified with the fundamental domain of the extended Weyl group
in $\gt$, which is given by the (solid) equilateral triangle
\begin{equation}\label{eq:clmcc}
  \eM_{cl}=\{ X\in\gt \mid 0 \leq \alpha_i(X) \leq 1~,\ i=1,2,3\}~.
\end{equation}
Notice that the boundary of $\eM_{cl}$ belongs to the singular lines
$L_{\alpha_i, n}$; in particular, its vertices belong to the
intersection of all three families of singular lines, and describe the
central elements of $\SU_3$.

\begin{figure}[h!]
  \setlength{\unitlength}{1mm}
  \begin{center}
    \subfigure[]{\label{fig:McSU3-a}
      \begin{picture}(50,40)(0,0)
        \put(10,1){%
          \put(0,0){\circle*{1}}
          \put(30,0){\circle*{1}}
          \put(15,25.9808){\circle*{1}}
          \thicklines
          \path(0,0)(30,0)(15,25.9808)(0,0)
          \thinlines
          \put(8,5){$\scriptstyle \SU_3/\U_1^2$}
          \put(0,20){\vector(1,-1){7.35}}
          \put(-8,23){$\frac{\SU_3}{\mathrm{S}(\U_2\times \U_1)}$}
          \put(24,34.9808){\vector(-1,-1){9}}
          \put(25,35){$\frac{\SU_3}{\SU_3}$}
          }
      \end{picture}
      }
    \qquad
    \subfigure[]{
      \begin{picture}(50,40)(0,0)
        \put(10,1){%
          \put(0,0){\circle*{1}}
          \put(30,0){\circle*{1}}
          \put(15,25.9808){\circle*{1}}
          \put(15,8.66025){\circle*{2}}
          \thicklines
          \path(0,0)(30,0)(15,25.9808)(0,0)
          \put(0,0){\color{light2}
            \path(24,0)(3,5.19615)(18,20.7846)(24,0)
            }
          \thinlines
          }
        \put(8,-2){$\scriptstyle 0$}
        \put(41,2){$\scriptstyle X_{\omega^2}$}
        \put(26,28){$\scriptstyle X_\omega$}
        \put(10,6.19615){$\scriptstyle X$}
        \put(29,21.7846){$\scriptstyle \omega\cdot X$}
        \put(32,-3){$\scriptstyle \omega^2\cdot X$}
      \end{picture}
      }
    \caption{Moduli space of conjugacy classes of $\SU_3$}
    \label{fig:McSU3}
  \end{center}
\end{figure}
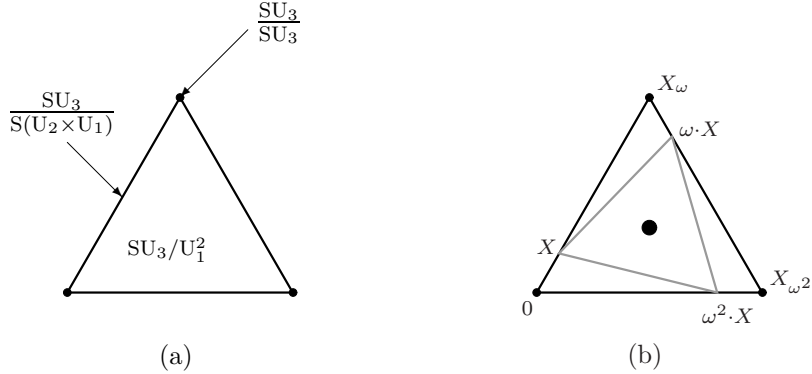

The description of a conjugacy class as a homogeneous space given by
\eqref{eq:cc} gives us a way to evaluate its dimension, as the
codimension of $\eC(h)$ is equal to the dimension of corresponding
centraliser $\eZ(h)$.  Clearly, if $h$ is regular---that is, if $h$ is 
contained in just one maximal torus---then the connected component of
$\eZ(h)$ agrees with the maximal torus and hence the dimension of
$\eZ(h)$ is equal to the rank of $\SU_3$.  On the other hand, if $h$
is singular that is, if $h$ is contained in more than one maximal
torus, then then dimension of the centraliser will be strictly larger
than the rank of the group.  In general, we thus have
\begin{equation*}
\dim\eC(h) \leq \dim\SU_3 - \rank\SU_3~.
\end{equation*}
The interior points in $\eM_{cl}$ are regular, and give rise to
$6$-dimensional conjugacy classes of the form $\SU_3/\U_1^2$.  If we
now consider an element $X$ in $\gt$ which belongs to one of the
singular lines, say, $L_{\alpha_i, n}$, this describes a singular
element $h=\exp(X)$ in $T$ whose centraliser includes the subgroup
$\SU_2^{\alpha_i}$.  Thus the boundary points belonging to the three
edges are singular, giving rise to $4$-dimensional conjugacy classes
of the form $\SU_3/\mathrm{S}(\U_2\times\U_1)$.  Finally, the three
vertices corresponding to the three central elements of $\SU_3$
describe point-like D-branes of the form $\SU_3/\SU_3$.  We thus
obtain the space of classical symmetric D-branes represented in
Figure~\ref{fig:McSU3-a}.

D-branes wrapping shifted conjugacy classes represent configurations
that are indistinguishable, in their intrinsic properties, from their
unshifted counterparts.  A brief inspection of the space $\eM_{cl}$
reveals that one has a natural action of the centre of $\SU_3$ on the
space of conjugacy classes, such that for any element $z$ in the
centre we can define a map
\begin{equation}\label{eq:equiv}
  \begin{aligned}[m]
    z : \eM_{cl} &\to \eM_{cl}\\
    \eC(X) &\mapsto \eC(z\cdot X)=z\eC(X)~,
  \end{aligned}
\end{equation}
which consists in shifting a given conjugacy class $\eC(X)$ by the
central element $z$.  If we denote the centre by
$\{e,\omega,\omega^2\}$, with $\omega^3=e$, we see that the orbit
$\eO(X)$ of a point $X$ in $\eM_{cl}$ under the above action of the
centre consists (with one exception, the point in the centre of the
equilateral triangle) of three points, $\eO(X)=\{X,\omega\cdot
X,\omega^2\cdot X\}$, which determine an equilateral triangle inside
$\eM_{cl}$ and concentric to it.  In particular, the three vertices of
$\eM_{cl}$ are one such orbit, $\eO(0)=\{0,X_\omega,X_{\omega^2}\}$,
where $\omega=\exp(X_\omega)$.  One can think of this action, which
shifts conjugacy classes, as defining a kind of equivalence relation
on the space of D-branes; this will turn out to be instrumental in
determining their charges (see Section~5).

\subsection{Quantum analysis}

The classical analysis of the possible D-brane configurations gives us
a continuous family of conjugacy classes \eqref{eq:clmcc},
parametrised by the points of a solid polygon in $\gt$.  In order to
determine which of these configurations are consistent at the quantum
level one has to analyse the quantisation conditions imposed by the
requirement of single-valuedness of the path integral
\cite{KlS,AS,Gaw,FSrc}.  This results in a number of quantisation
conditions corresponding to evaluating the global worldsheet anomaly
on the $3$-cycles in $H_3(\SU_3,\eC)$.

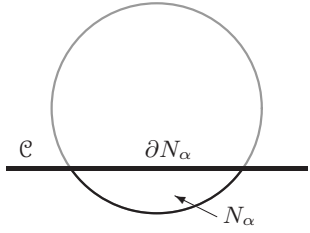
\begin{figure}[h!]
  \setlength{\unitlength}{0.8mm}
  \begin{center}
    \begin{picture}(50,50)
      \thinlines
      \put(35,7){\vector(-2,1){7}}
      \put(36,6){\scriptsize{$N_\alpha$}}
      \thicklines
      \put(0,0){\color{light2}
        \put(25,25){\circle{35}}
        }
      \put(25,25){\arc{35}{0.608246}{2.53335}}
      \Thicklines
      \put(0,15){\line(1,0){50}}
      \put(2,17){\scriptsize{$\eC$}}
      \put(23,17){\scriptsize{$\partial N_\alpha$}}
    \end{picture}
  \end{center}
  \caption{Relative cycle in $H_3(\SU_3,\eC)$}
  \label{fig:relcycle}
\end{figure}

The data necessary for implementing this requirement is the relative
$3$-cocycle $(H,\omega)$ in $H^3(\SU_3,\eC)$ and a basis of relative
cycles in $H_3(\SU_3,\eC)$.  The relative form $(H,\omega)$ is well
known: $H$ is the standard three-form field on the $\SU_3$ and
$\omega$ is the two-form field \cite{AS,Gaw,Q0} defined on the the
D-brane which satisfies $d\omega = H|_{\eC}$.  The other piece of
data, the basis of relative $3$-cycles, is provided by the
$\SU_2^\alpha$ subgroups of $\SU_3$.  In order to be more precise, let
us fix a given conjugacy class $\eC (h)$, with $h=\exp(X)$, for $X$ in
$\gt$; then for every root $\alpha$ we can exhibit a relative cycle
$(N_\alpha,\d N_\alpha)$ in $H_3(\SU_3,\eC)$, schematically
represented in Figure~\ref{fig:relcycle}.  To this end we denote by
$\{E_{\pm\alpha},H_\alpha\}$ the standard basis of the corresponding
$\su(2)_\alpha$.  If we split $X = X^\perp + X_\alpha$, where
$X_\alpha = \half\alpha(X)H_\alpha$, we can write $h$ as a product, $h
= h^\perp h_\alpha$, where $h_\alpha = \exp(X_\alpha)$ belongs to the
$\SU_2^\alpha$ subgroup, whereas $h^\perp = \exp(X^\perp)$ commutes
with it.  Using this, one can deduce that our conjugacy class $\eC(h)$
intersects the coset $h^\perp\SU_2^\alpha$ exactly in a conjugacy
class of $\SU_2^\alpha$, shifted in $\SU_3$:
\begin{equation}\label{eq:intersection}
\eC(h;\SU_3) \cap h^\perp\SU_2^\alpha =
                                   h^\perp\eC(h_\alpha;\SU_2)~,
\end{equation}
where we have explicitly indicated to which group each of the two
conjugacy classes belong.  

This allows us to define the relative cycle $(N_\alpha,\d N_\alpha)$
by taking $N_\alpha$ to be a $3$-submanifold of $h^\perp
\SU_2^\alpha$, whose boundary $\d N_\alpha$ is given by the shifted 
conjugacy class above:
\begin{equation}\label{eq:2cycle}
\d N_\alpha = h^\perp \eC(h_\alpha;\SU_2)~.
\end{equation}
Clearly, from \eqref{eq:intersection} we have that $\d N_\alpha
\subset\eC(X)$.  Now it is easy to see that the global worldsheet
anomaly evaluated for this particular relative $3$-cycle of $\SU_3$ is
proportional to the one computed for the case of $\SU_2$ and the above
two-dimensional conjugacy class.  Indeed, one obtains
\begin{equation}\label{eq:qcalpha}
\frac{1}{2\pi}\left(\int_{N_\alpha} H - \int_{\d N_\alpha}
             \omega\right) = k\alpha(X)~,
\end{equation}
since induced theory on $\SU_2$ has level $k$ as well.  Hence the
requirement that the path integral be single-valued forces
$k\alpha(X)$ to take integral values, for all roots of $\SU_3$.
Moreover, we have an additional quantisation condition which
corresponds to the honest $3$-cycle in $H_3(\SU_3,\eC)$,
\begin{equation*}
\frac{1}{2\pi}\int_{\SU_2^\alpha} H = k~,
\end{equation*}
which is nothing but the quantisation of the level, familiar from the
standard WZW theory.

We have thus obtained the space $\eM_q:=\eM_q(\SU_3,\1)$ of
symmetric D-branes in $\SU_3$ at level $k$
\begin{equation}\label{eq:qms1}
  \eM_q = \left\{X\in\gh \mid k\alpha_i(X)\in\Z,\ 0
               \leq k\alpha_i(X)\leq k,\ i=1,2,3 \right\}~. 
\end{equation}
Alternatively, we can use the isomorphism $\gt\cong\gt^*$ to associate
to every $X$ in $\gt$ an element $\lambda$ in $\gt^*$, in terms of
which the space of D-brane configurations becomes \cite{Gaw}
\begin{equation*}
\eM_q = \{\lambda\in\gh^* \mid \la\alpha_i,\lambda\ra\in\Z,\ 
               0 \leq\la\alpha_i,\lambda\ra \leq k,\ i=1,2,3 \}~.
\end{equation*}
This shows that the set of consistent symmetric D-brane configurations
in $\SU_3$ at level $k$ is in one-to-one correspondence with the set
of IHW representations of the affine Lie algebra
$\widehat\su(3)^{(1)}_k$.

\begin{figure}[h!]
  \setlength{\unitlength}{0.9mm}
  \begin{center}
    \begin{picture}(130,40)
      \put(5,10){%
        \put(0,0){\circle*{2}}
        \put(30,0){\circle*{2}}
        \put(15,25.9808){\circle*{2}}
        \put(10,-8){\makebox{$k=1$}}
        }
      \put(50,10){%
        \put(0,0){\circle*{2}}
        \put(15, 0){\circle*{2}}
        \put(7.5, 12.9904){\circle*{2}}
        \put(22.5, 12.9904){\circle*{2}}
        \put(30,0){\circle*{2}}
        \put(15,25.9808){\circle*{2}}
        \put(10,-8){\makebox{$k=2$}}
        }
      \put(95,10){%
        \put(0,0){\circle*{2}}
        \put(10,0){\circle*{2}}
        \put(20,0){\circle*{2}}
        \put(30,0){\circle*{2}}
        \put(5, 8.66025){\circle*{2}}
        \put(10, 17.3205){\circle*{2}}
        \put(15, 8.66025){\circle*{2}}
        \put(20, 17.3205){\circle*{2}}
        \put(25, 8.66025){\circle*{2}}
        \put(15,25.9808){\circle*{2}}
        \put(10,-8){\makebox{$k=3$}}
        }
    \end{picture}
    \caption{Quantum moduli space for $\SU_3$ for lowest values of
    the level $k$.}
    \label{fig:MqSU3}    
  \end{center}
\end{figure}
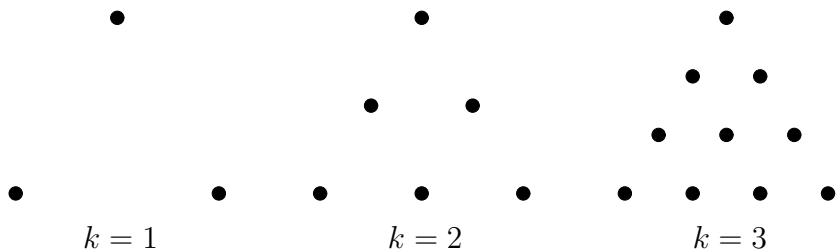

The space of symmetric D-brane configurations in $\SU_3$ for the first
few values of the level $k$ is represented in Figure~\ref{fig:MqSU3}.
At a given level $k$ we have $3$ point-like, $3(k-1)$ $4$-dimensional
and $\half(k-1)(k-2)$ $6$-dimensional symmetric D-branes.  We also see
that the $4$-dimensional conjugacy classes are characterised by
quantum numbers $(\lambda_1,\lambda_2)$ with either one of the
$\lambda$'s being equal to zero or $\lambda_1+\lambda_2 = k$; the
point-like conjugacy classes are described by $(0,0)$, $(0,k)$,
$(k,0)$.  In particular, the lower-dimensional conjugacy classes
dominate the spectrum of D-branes for $k\leq 9$.

Finally, notice that the action of the centre on $\eM_{cl}$ does
induce a similar action on the discrete set of states $\eM_q$ which
results in the various D-brane configurations in $\eM_q$ being
organised into triplets, the three elements of a triplet differing
from one another just by a translation by a central element in
$\SU_3$.  From an algebraic point of view, this corresponds to the
action of the extended Dynkin diagram on the IHW representations of
the affine Lie algebra, as was recently pointed out in \cite{MMS2}.

\section{$\SO_3$ versus $\SU_2$}

We pause for a moment our analysis of D-branes in $\SU_3$ in order to
discuss the possible D-submanifolds in the non-simply connected group
$\SO_3$, which will prove instrumental in our discussion of twisted
branes in the next section.

\begin{figure}[h!]
  \begin{center}
    \begin{picture}(90,7)(15,0)
      \multiput(15,1)(30,0){4}{\circle{2}}
      \multiput(30,1)(30,0){3}{\circle*{2}}
      \multiputlist(7.5,1)(15,0)%
      {$\cdots$,{\line(1,0){13}},{\line(1,0){13}},{\line(1,0){13}},%
        {\line(1,0){13}},{\line(1,0){13}},{\line(1,0){13}},$\cdots$}
      \multiputlist(15,-2)(15,0){$\scriptstyle L_{\alpha,-3}$,%
        $\scriptstyle L_{\alpha,-2}$,$\scriptstyle L_{\alpha,-1}$,%
        $\scriptstyle L_{\alpha,0}$,$\scriptstyle L_{\alpha,1}$,%
        $\scriptstyle L_{\alpha,2}$,$\scriptstyle L_{\alpha,3}$}
    \end{picture}
    \vspace{8pt}
    \caption{Stiefel diagram of $\SU_2$.  Circles indicate the
    central lattice $\Lambda_Z(\SU_2)$ and filled circles indicate
    the integer lattice $\Lambda_I(\SU_2)$.}
    \label{fig:SSU2}
  \end{center}
\end{figure}
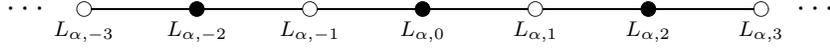

\begin{figure}[h!]
  \setlength{\unitlength}{1mm}
  \begin{center}
  \mbox{%
    \subfigure{
    \begin{picture}(40,30)
      \Thicklines
      \put(10.5,15){\line(1,0){19}}
      \thinlines
      \put(30,15){\shade\circle{1}}
      \put(10,15){\circle*{1}}
      \put(20,20){\vector(0,-1){4.5}}
      \put(16,23){$\frac{\SU_2}{\U_1}$}
      \put(10,10){\vector(0,1){4.3}}
      \put(6,5){$\frac{\SU_2}{\SU_2}$}
      \put(30,10){\vector(0,1){4.3}}
      \put(26,5){$\frac{\SU_2}{\SU_2}$}
    \end{picture}
    }
  \qquad
  \subfigure{
    \begin{picture}(40,30)
      \Thicklines
      \put(10.5,15){\line(1,0){19}}
      \thinlines
      \put(30,15){\shade\circle{1}}
      \put(10,15){\circle*{1}}
      \put(20,20){\vector(0,-1){4.5}}
      \put(16,23){$\frac{\SO_3}{\SO_2}$}
      \put(10,10){\vector(0,1){4.3}}
      \put(6,5){$\frac{\SO_3}{\SO_3}$}
      \put(30,10){\vector(0,1){4.3}}
      \put(26,5){$\frac{\SO_3}{\O_2}$}
    \end{picture}
}}
    \caption{Moduli space of conjugacy classes of $\SU_2$ and $\SO_3$}
    \label{fig:McSU2}
  \end{center}
\end{figure}
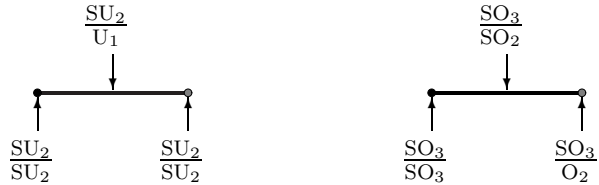

The groups $\SU_2$ and its quotient $\SO_3$ only admit standard
conjugacy classes as they do not possess outer automorphisms.  The
moduli spaces of symmetric D-branes in $\SU_2$ is well understood
\cite{AS}, and we briefly describe it here only for later comparison.
The corresponding Stiefel diagram is represented in
Figure~\ref{fig:SSU2}.  The classical moduli space, represented in
Figure~\ref{fig:McSU2}, is given by the fundamental domain of the
extended Weyl group in the Cartan subalgebra $\gt$:
\begin{equation*}
\eM_{cl}(\SU_2)=\{ X\in\gt \mid 0\leq\alpha(X)\leq 1\}~,
\end{equation*}
The two extremities correspond to the point-like D-branes, which are
described as $\SU_2/\SU_2$, whereas the interior points describe
regular elements in $\SU_2$ which give rise to $2$-dimensional D-branes
of the form $\SU_2/\U(1)$.

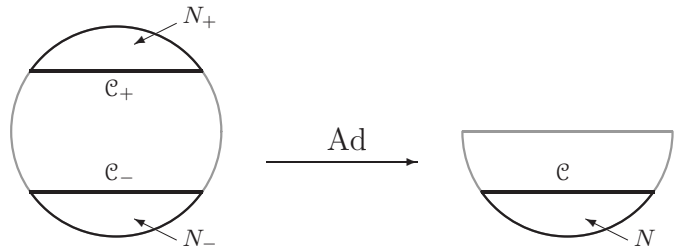
\begin{figure}[h!]
  \setlength{\unitlength}{0.8mm}
  \begin{center}
    \begin{picture}(125,50)
      \thinlines
      \put(35,7){\vector(-2,1){7}}
      \put(36,6){\scriptsize{$N_-$}}
      \put(35,43){\vector(-2,-1){7}}
      \put(36,43){\scriptsize{$N_+$}}
      \thicklines
      \put(0,0){\color{light2}
        \put(25,25){\circle{35}}
        }
      \put(25,25){\arc{35}{0.608246}{2.53335}}
      \put(25,25){\arc{35}{3.74984}{5.67494}}
      \Thicklines
      \put(10.6386,15){\line(1,0){28.7228}}
      \put(10.6386,35){\line(1,0){28.7228}}
      \put(23,17){\scriptsize{$\eC_-$}}
      \put(23,31){\scriptsize{$\eC_+$}}
      \thinlines
      \put(50,20){\vector(1,0){25}}
      \put(60,22){$\Ad$}
      \thinlines
      \put(75,0){
        \put(35,7){\vector(-2,1){7}}
        \put(36,6){\scriptsize{$N$}}
        \thicklines
        \put(0,0){\color{light2}
          \put(25,25){\arc{35}{0}{3.14159}}
          \put(7.5,25){\line(1,0){35}}
          }
        \put(25,25){\arc{35}{0.608246}{2.53335}}
        \Thicklines
        \put(10.6386,15){\line(1,0){28.7228}}
        \put(23,17){\scriptsize{$\eC$}}
        }
    \end{picture}
  \end{center}
  \caption{Relative $3$-cycles in $\SU_2$ and $\SO_3$.}
  \label{fig:relcyclesSU2SO3}
\end{figure}

The requirement of single-valuedness of the path integral for the
$\SU_2$ WZW model imposes two quantisation conditions, corresponding
to the two generating cycles of $H_3(\SU_2,\eC)$, which are the
$\SU_2$ manifold itself and the relative $3$-cycle $(N_-,\d N_-)$
with $\d N_- = \eC_-(X)$ (see Figure~\ref{fig:relcyclesSU2SO3}).  The first
quantisation condition, corresponding to the honest $3$-cycle,
requires that the following
\begin{equation*}
\frac{1}{2\pi}\int_{\SU_2} H_{\SU_2} = k~,
\end{equation*}
take integer values, which essentially reiterates the quantisation of
the level, $k\in\Z$.  The second quantisation condition, which imposes
that 
\begin{equation*}
\frac{1}{2\pi}\left(\int_{N_-} H_{\SU_2} - \int_{\d N_-}
                          \omega_{\SU_2}\right) = k\alpha(X)~,
\end{equation*}
take integer values, selects a discrete set of conjugacy classes,
which are labeled by the IHW representations of the affine Lie algebra
$\widehat\su(2)^{(1)}_k$.  Thus the space of consistent D-brane
configurations is given by
\begin{equation*}
\eM_{q}(\SU_2) = \{ X\in\gt \mid k\alpha(X)\in\Z~,\
                  0\leq k\alpha(X)\leq k\}~, 
\end{equation*}

The non-simply connected group $\SO_3$ can be obtained from $\SU_2$
by factoring out its centre: $\SO_3\cong\SU_2/\Z_2$.  Let $\Ad$ be
the group homomorphism
\begin{equation*}
\Ad:\SU_2 \to \SO_3,
\end{equation*}
which projects the central subgroup $\Z_2$ of $\SU_2$ onto the
identity element in $\SO_3$.  Using this projection, one can easily see
that any conjugacy class $\eC(h)$ in $\SO_3$ is the image, under $\Ad$,
of a conjugacy class in the simply connected $\SU_2$:
\begin{equation*}
\eC(h) = \Ad\eC(\Ad^{-1}(h))~.
\end{equation*}
Notice that $\Ad^{-1}(h)$ consists of two points in $\SU_2$, which
means that, generically, $\Ad$ projects two conjugacy classes in
$\SU_2$ to one conjugacy class in $\SO_3$.  (The exception is
provided by the `equatorial' conjugacy class in $\SU_2$.)

The groups $\SU_2$ and $\SO_3$ share the same root system, which
implies that they also have the same Stiefel diagram and, in
particular, the same fundamental domain of the extended Weyl group.
Since the integral lattice of $\SO_3$ is identical to its central
lattice, in this case in the Stiefel diagram shown in
Figure~\ref{fig:SSU2} all the circles are filled.  Nevertheless, in
the case of $\SO_3$, the fundamental domain of the extended Weyl
group can no longer be identified with the moduli space of symmetric
D-branes; instead, the latter is obtained by a $\Z_2$ quotient of the
former
\begin{equation*}
\eM_{cl}(\SO_3) = \{ X\in\gt \mid 0\leq\alpha(X)\leq \half\}~.
\end{equation*}
Here the point $\alpha(X)=0$ describes a singular point (the identity
in $\SO_3$) and a point-like D-brane $\SO_3/\SO_3$.  The other
extremity $\alpha(X)=\half$ is a regular point with a non-connected
centraliser, and the corresponding conjugacy class is a non-orientable
submanifold $\SO_3/\O_2$.  As before, the interior points are
regular, and the corresponding symmetric D-branes wrap the homogeneous
spaces $\SO_3/\SO_2$.  This is illustrated in
Figure~\ref{fig:McSU2}.

The quantisation conditions for $\SO_3$ can be analysed using the
$\SU_2$ results.  Let us start with the `generic' case $\eC \cong
\SO_3/\SO_2$.  The requirement of single-valuedness of the path
integral for the $\SO_3$ WZW model imposes two quantisation
conditions, corresponding to the two generating cycles of
$H_3(\SO_3,\eC) \cong \Z\oplus\Z$.  These relative cycles can be obtained
from the $\SU_2$ relative cycles, via the projection homomorphism
$\Ad$, as is schematically indicated in
Figure~\ref{fig:relcyclesSU2SO3}.  Indeed, the generic relative cycle
in $\SU_2$, given by $(N_-,\d N_-)$, can be mapped into the relative
cycle $(N,\d N)$ of $\SO_3$, where $\Ad(N_-) = N$ and $\Ad(\eC_-) =
\eC$.  By contrast, since $\SU_2$ covers $\SO_3$ twice, we obtain
$\Ad_*[\SU_2] = 2[\SO_3]$ in homology.

The first quantisation condition, corresponding to the honest
$3$-cycle, requires that the following be integer valued:
\begin{equation*}
\frac{1}{2\pi}\int_{\SO_3} H_{\SO_3} = \frac{1}{4\pi}\int_{\SU_2}
H_{\SU_2} = \frac{k}{2}~,
\end{equation*}
where we used $\Ad^* H_{\SO_3} = H_{\SU_2}$.  This recovers the well
known fact that the level in the $\SO_3$ theory has to be even, so
that $k\in 2\Z$.  On the other hand, a similar argument, employing
$\Ad(N_-,\d N_-) = (N,\d N)$ and $\Ad^* (H_{\SO_3},\omega_{\SO_3}) =
(H_{\SU_2},\omega_{\SU_2})$, shows that the second quantisation
condition for the $\SO_3$ theory is identical to the corresponding one
in the case of $\SU_2$
\begin{equation*}
\frac{1}{2\pi}\left(\int_{N} H_{\SO_3} - \int_{\d N}
           \omega_{\SO_3}\right) = k\alpha(X)~.
\end{equation*}
This selects a discrete set of consistent D-brane configurations which
can also be labeled by the IHW representations of the affine Lie
algebra $\widehat\su(2)^{(1)}_k$.

For the two conjugacy classes at the boundary of the moduli space that
is, the point-like $\SO_3/\SO_3$ and the non-orientable
$\SO_3/\O_2$, we have $H_3(\SO_3,\eC) \cong \Z$, thus the only
quantisation condition is the one corresponding to the honest
$3$-cycle, which we know to be satisfied.

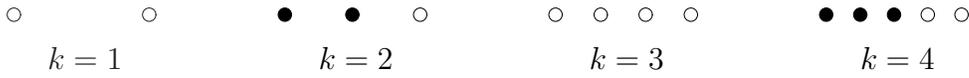
\begin{figure}[h!]
  \setlength{\unitlength}{0.9mm}
  \begin{center}
    \begin{picture}(150,20)
      \put(5,10){%
        \put(0,0){\circle{2}}
        \put(20,0){\circle{2}}
        \put(5,-8){\makebox{$k=1$}}
        }
      \put(45,10){%
        \put(0,0){\circle*{2}}
        \put(10, 0){\circle*{2}}
        \put(20,0){\circle{2}}
        \put(5,-8){\makebox{$k=2$}}
        }
      \put(85,10){%
        \put(0,0){\circle{2}}
        \put(6.6667,0){\circle{2}}
        \put(13.3333,0){\circle{2}}
        \put(20,0){\circle{2}}
        \put(5,-8){\makebox{$k=3$}}
        }
      \put(125,10){%
        \put(0,0){\circle*{2}}
        \put(5,0){\circle*{2}}
        \put(10,0){\circle*{2}}
        \put(15,0){\circle{2}}
        \put(20,0){\circle{2}}
        \put(5,-8){\makebox{$k=4$}}
        }
    \end{picture}
    \caption{Quantum moduli spaces for $\SO_3$ (black circles) and
      $\SU_2$ for lowest values of the level $k$.}
    \label{fig:MqSU2SO3}    
  \end{center}
\end{figure}

Therefore, by contrast with $\SU_2$, in this case we only get $\frac
k2 + 1$ admissible D-submanifolds at a given level $k$, these being
described by
\begin{equation*}
\eM_{q}(\SO_3) = \left\{ X\in\gt \mid k\alpha(X)\in\Z~,\
                  0\leq k\alpha(X)\leq \frac{k}{2} \right\}~, 
\end{equation*}
which is exemplified for the first few levels in
Figure~\ref{fig:MqSU2SO3}.

\section{Twisted D-branes in $\SU_3$}

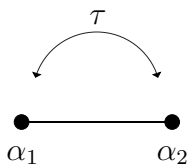
\begin{figure}[h!]
  \begin{center}
    \begin{picture}(40,25)
      \thinlines
      \put(20,11.47774){\arc{17.0453}{3.44209}{5.98269}}
      \blacken\path(11.5297,14.6216)(11.8593,14)(12.4849,14.3296)(11.5297,14.6216)
      \blacken\path(28.4703,14.6216)(28.1407,14)(27.5151,14.3296)(28.4703,14.6216)
      \put(10.5,8){\line(1,0){19}}
      \put(30,8){\circle*{2}}
      \put(10,8){\circle*{2}}
      \put(8,3){$\alpha_1$}
      \put(28,3){$\alpha_2$}
      \put(19,21){$\tau$}
    \end{picture}
    \caption{Dynkin diagram automorphism of $\SU_3$}
    \label{fig:DdSU3}
  \end{center}
\end{figure}

The group of (metric-preserving) outer automorphisms can be defined as
the factor group
\begin{equation*}
\Out_o (G) = \Aut_o (G)/\Inn_o (G)~,
\end{equation*}
where $\Aut_o(G)$ denotes the group of metric-preserving automorphisms
of $G$ and $\Inn_o(G)\subset \Aut_o(G)$ denotes the invariant subgroup
corresponding to inner automorphisms.\footnote{It was suggested in
  \cite{FSNW} that this group plays a role analogous to that of the
  T-duality group in toroidal compactifications.}  In the particular
case of $\SU_3$ this group contains only one non-trivial element.  One
particular representative for this outer automorphism is given by
complex conjugation.  For the purpose of our analysis we will however
take the outer automorphism to be the Dynkin diagram automorphism
$\tau$ illustrated in Figure~\ref{fig:DdSU3}, which acts on the simple
roots of $\SU_3$ by interchanging $\alpha_1$ and $\alpha_2$.  This
type of automorphisms have been studied in some detail in the
mathematics literature, since they are central to the construction of
twisted affine Lie algebras (see, for instance, \cite{Kac}). Notice
that $\tau$ has order two, that is, $\tau^2=\1$.

The diagram automorphism $\tau$ gives rise to a `folded' root system,
denoted by $\Phi^\tau$, which is generated by the linear combinations
of the roots of $\su(3)$ that are invariant under $\tau$:
\begin{equation}\label{eq:baralpha}
\bar\alpha = \begin{cases}
             \alpha~, &\text{for}\ \tau(\alpha)=\alpha\\
             \half (\alpha + \tau(\alpha))~, &\text{for}\ \tau(\alpha) \neq
             \alpha
             \end{cases}
\end{equation}
We thus have that the set of positive roots is given by
$\Phi^\tau_+=\{(\alpha_1+\alpha_2)/2,\alpha_1 + \alpha_2\}$.  Notice
that the folded root system $\Phi^\tau$ is not a subset of the
original root system $\Phi$.  In fact, in this case, $\Phi^\tau=BC_1$
is not even the root system of a Lie algebra.

This automorphism, defined at the level of $\Delta$, can be extended
by linearity to $\gt^*$, and further, by duality, to the Cartan
subalgebra, in such a way that for any $X$ in $\gt$
\begin{equation*}
\tau(\alpha) (\tau(X)) = \alpha(X)~.
\end{equation*}
If we then define
\begin{equation*}
\tau(E_\alpha) = E_{\tau(\alpha)}~, \qquad\qquad \alpha\in\Delta~,
\end{equation*}
one extends $\tau$ to a Lie algebra automorphism.  Furthermore, since
$\SU_3$ is compact, connected, and simply connected, we can lift
$\tau$ to a Lie group automorphism which, by a small abuse of
notation, we will also denote by $\tau$.  It is known that there
exists a maximal torus $T$ which is left invariant by $\tau$.  If we
denote by $\SU_3^\tau$ the fixed point subgroup of $\SU_3$ under
$\tau$, and by $T^\tau$ the fixed point set of $T$, we clearly have
that $T^\tau$ is the maximal torus of $\SU_3^\tau$ and
$\tau|_{\SU_3^\tau} = \1$.  Let $\su(3)^\tau$ and $\gt^\tau$ be the
Lie algebras of $\SU_3^\tau$ and $T^\tau$ respectively; these are
nothing but the fixed point sets of $\tau$ in $\su(3)$ and $\gt$,
respectively.  The root system of $\SU_3^\tau$ has one generator
$\bar\alpha$ which is given by
\begin{equation}\label{eq:alphabar}
\bar\alpha = \frac{\alpha_1 + \alpha_2}{2}~.
\end{equation}

One can determine the fixed point subgroup $\SU_3^\tau$ in a variety
of ways (see, for instance, \cite{Wendt}) and
obtain\footnote{Alternatively, one can choose to work with a different
  representative $\rho$ of the outer automorphism of $\SU_3$ which is
  given by complex conjugation, $\rho(g)=\bar g$.  The resulting fixed 
  point subgroup $\SU_3^\rho=\SO_3$ will then be conjugate and hence
  isomorphic to $\SU_3^\tau$.}
\begin{equation*}
\SU_3^\tau \cong \SO_3~,
\end{equation*}
with the simple root $\bar\alpha$ given by \eqref{eq:alphabar}.  The
coroot lattice $\Lambda^\vee_{\bar R}$ of $\SU_3^\tau$ acts on
$\gt^\tau$ by translations, and is generated by
\begin{equation}\label{eqref:coroot}
\bar\alpha^* = 2(\alpha_1^* + \alpha_2^*)~.
\end{equation}

\subsection{Twisted D-branes}

The twisted conjugacy class $\eC_\tau(g)$ of a group element $g$ can
be defined as the orbit of $g$ under the twisted conjugation
$\Ad_{\tau,k}:g\mapsto\tau(k)gk^{-1}$, for any $k$ in the group
manifold.  Similarly to the untwisted case, this takes the form of a
homogeneous space
\begin{equation*}
\eC_\tau(g) \cong G/\eZ_\tau(g)~,
\end{equation*}
where $\eZ_\tau(g)$ denotes the twisted centraliser of $g$ defined as
the subgroup
\begin{equation*}
\eZ_\tau(g) = \{k\in G \mid \tau(k)g = gk\}~.
\end{equation*}

Let $\Gamma=\{\1,\tau\}$ be the finite group generated by the diagram
automorphism and let
\begin{equation*}
\eG = \Gamma \ltimes \SU_3
\end{equation*}
be the \emph{principal extension} of $\SU_3$ by $\tau$.  It can be
characterised as the smallest Lie group containing $\SU_3$ for which
$\tau$ is an inner automorphism.  It is the (disjoint) union of two
connected components, which we write symbolically as $\eG = \SU_3 +
\tau\SU_3$, with $\SU_3$ being the connected component of the
identity.  $\SU_3$ acts on $\eG$ by conjugation, and since $\SU_3$ is
connected, this action stabilises each of the connected components of
$\eG$.  On $\SU_3$ it is just the standard conjugation of $\SU_3$ on
itself; but on $\tau \SU_3$ it is the twisted conjugation defined
above (provided that $\tau^2 = \1$).  It then follows that the
conjugacy class $\eO(\tau g)$ of an element $\tau g$ under $\SU_3$ is
nothing but the twisted conjugacy class shifted by $\tau$; that is,
\begin{equation}\label{eq:occ}
  \eO(\tau g) = \tau\eC_\tau(g)~.
\end{equation}
This relation is clearly valid for any Lie group $G$, provided
$\tau^2=\1$; the connection between adjoint orbits in $G\tau$ and
twisted adjoint orbits in $G$ allows us to use the theory of
nonconnected Lie groups to to give a detailed description of the space
of twisted conjugacy classes.

One of the key notions which we will need to use in the next paragraph
is that of $\tau$-regularity, which can be thought of as a
translation, at the level of $G$, of the notion of regularity in $\tau
G$.  An element $\tau g$ is said to be regular in $\tau G$ if it
belongs to only one Cartan subgroup $S$ of $\eG$
\cite{BroeckerTomDieck}.  Alternatively, we can say that $\tau g$ is
regular in $\tau G$ if the connected component of the identity of its
centraliser $\eZ(\tau g)$ is abelian.  Using the relation between
$\eZ(\tau g)$ and $\eZ_\tau(g)$, we say that an element $g$ in $G$ is
\emph{$\tau$-regular} if the connected component of its twisted
centraliser is abelian.  Otherwise, $g$ is said to be
\emph{$\tau$-singular} in $G$.

\subsection{Classical analysis}

Using the theory on non-connected Lie groups, one can show
\cite{Siebenthal, Wendt, SD2notes} that twisted conjugacy classes in a
group $G$ are parametrised by the quotient
\begin{equation*}
\eM_{cl}(G,\tau) := T(G^\tau)/W_\tau(G)~,
\end{equation*}
where the `twisted' Weyl group $W_\tau(G)$ is given \cite{Wendt} by 
the semidirect product
\begin{equation*}
  W_\tau(G) = W(G^\tau) \ltimes \Lambda_T~,
\end{equation*}
of the Weyl group of $G^\tau$ with a certain discrete group,
$\Lambda_T = \left(T/T^\tau\right)^\tau$, which acts on $\gt^\tau$ by
translations.  The generators $\alpha_T$ of this translation group
stem from those coroots $\alpha^*$ of $G$ for which
$\tau(\alpha^*)\neq\alpha^*$, and are given by $\tau$-invariant linear
combinations of these:
\begin{equation*}
\alpha^*_T = \frac{\alpha^* + \tau(\alpha^*)}{2}~.
\end{equation*}

The space of twisted conjugacy classes in $\SU_3$ is therefore
described by the quotient
\begin{equation*}
\eM_{cl}(\SU_3,\tau) = \frac{\gt^\tau}{W(\SO_3) \ltimes \left(
                    \Lambda_I(\SO_3) \times \Lambda_T\right)}~.
\end{equation*}
Notice that, in the absence of $\Lambda_T$, the above quotient would
be nothing but the space of standard conjugacy classes in the fixed
point subgroup $\SU_3^\tau\cong\SO_3$.  In order to fully understand
the structure of this space and thus the spectrum of possible twisted
D-branes in $\SU_3$, we need to understand better the twisted Weyl
group $W_\tau(\SU_3)$.  This is generated by the Weyl reflection
\begin{equation*}
s_{\bar\alpha}: X \mapsto - X~,
\end{equation*}
and by the translation in $\Lambda_T$
\begin{equation*}
\gamma_{\alpha_T}: X \mapsto X + \alpha_T^*~,\qquad\qquad 
\alpha_T^* = \frac{1}{4} \bar\alpha^*~.
\end{equation*}
To these we have to add of course the translations corresponding to
the integral lattice of $\SO_3$.  

What is the effect of the `special' twisted Weyl transformations in
$W_\tau(G)$ which are generated by the translations in $\Lambda_T$?
It turns out that these translations result in some of the regular
points in $T^\tau$ becoming singular with respect to the twisted
adjoint action.  In other words, the effect of the translations in
$\Lambda_T$ amounts to having an additional number of singular lines
in $\gt^\tau$.  These new singular lines can be thought of as being
generated by the element in $(\gt^\tau)^*$ dual to $\alpha_T^*$:
\begin{equation}\label{eqref:alphaT}
\alpha_T = 2(\alpha + \tau(\alpha))~.
\end{equation}
We can thus construct a figure that could be called the `twisted
Stiefel diagram' of $\SU_3$, which is a diagram in $\gt^\tau$ defined 
as the inverse image, under the exponential map, of the $\tau$-singular
points in $T^\tau$.  This consists of two types of $0$-dimensional
singular hyperplanes: 
\begin{align*}
L_{\bar\alpha,n}~:\qquad &\bar\alpha(X)=n~,\\
S_{\alpha_T,n}~:\qquad &\alpha_T(X) = n~,\qquad \text{with}\quad
                                          \alpha_T=4\bar\alpha~. 
\end{align*}

The `ordinary' singular hyperplanes $\{L_{\bar\alpha,n}\}_{n\in\Z}$ do
not require much explanation.  They are generated by the $\SO_3$ root
$\bar\alpha$, and describe the singular points in $T^\tau$, which give
rise to lower-dimensional standard conjugacy classes in $\SO_3$; the
points belonging to these lines are, at the same time, $\tau$-singular
in $\SU_3$, since $W(\SO_3)\subset W_\tau(\SU_3)$.  By contrast
with the standard case, we have an additional family
$\{S_{\alpha_T,n}\}_{n\in\Z}$ of singular points, which describe the
elements in $T^\tau$ which are left invariant by a twisted Weyl
transformation.  Each of these singular points $X$ in $\gt^\tau$,
although might be regular in $\SO_3$, turn out to be $\tau$-singular
in $\SU_3$, giving rise to lower-dimensional twisted conjugacy
classes.

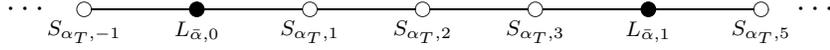
\begin{figure}[h!]
  \begin{center}
    \begin{picture}(90,7)(15,0)
      \multiput(15,1)(90,0){2}{\circle{2}}
      \multiput(45,1)(15,0){3}{\circle{2}}
      \multiput(30,1)(60,0){2}{\circle*{2}}
      \multiputlist(7.5,1)(15,0)%
      {$\cdots$,{\line(1,0){13}},{\line(1,0){13}},{\line(1,0){13}},%
        {\line(1,0){13}},{\line(1,0){13}},{\line(1,0){13}},$\cdots$}
      \multiputlist(15,-2)(15,0){$\scriptstyle S_{\alpha_T,-1}$,%
        $\scriptstyle L_{\bar\alpha,0}$,$\scriptstyle S_{\alpha_T,1}$,%
        $\scriptstyle S_{\alpha_T,2}$,$\scriptstyle S_{\alpha_T,3}$,%
        $\scriptstyle L_{\bar\alpha,1}$,$\scriptstyle S_{\alpha_T,5}$}
    \end{picture}
    \vspace{8pt}
    \caption{`Twisted' Stiefel diagram of $\SU_3$.  Circles indicate
      $\tau$-singular points and filled circles indicate singular
      points in the maximal torus of $\SU_3^\tau$.}
    \label{fig:tSSU3}
  \end{center}
\end{figure}

If the inverse image of $h$, under the exponential map, does not
belong to one of the special singular hyperplanes, then we have 
\begin{equation*}
T_h\eZ_\tau (h,\SU_3) = T_h\eZ (h,\SO_3)~,
\end{equation*}
from which we immediately deduce that
\begin{equation}\label{eq:codim}
\codim_{\SU_3}\eC_\tau(h) = \codim_{\SO_3} \eC(h)~.
\end{equation}

Let us now show that the points belonging to $S_{\alpha_T,n}$ are
indeed $\tau$-singular.  Notice first of all that
$S_{\alpha_T,4n}=L_{\bar\alpha,n}$, and the corresponding points are
singular already in $\SO_3$.  Furthermore, every point on
$S_{\alpha_T,4n+2}$ or $S_{\alpha_T,2n+1}$ is related, via a 
twisted Weyl transformation, to a point on $S_{\alpha_T,2}$ or
$S_{\alpha_T,1}$, respectively.  Hence it is sufficient to discuss
these latter two cases.

Let us start with the point $X$ on $S_{\alpha_T,2}$.  The corresponding
group element $h$ is regular in $\SO_3$, since $X$ lies inside an
alcove of $\SO_3$; however, since $X$ is conjugate, via a twisted
Weyl transformation (a $\Lambda_T$ translation, to be more precise),
to $X=0$, which is the point on $S_{\alpha_T,0}=L_{\bar\alpha,0}$, we
deduce that $h$ is $\tau$-singular in $\SU_3$, as its twisted
centraliser is $\SO_3$.

Finally, let us consider the case where $X$ belongs to
$S_{\alpha_T,1}$; this means that $X$ satisfies $2(\alpha_1 +
\alpha_2)(X)=1$; it also means that $X$ is regular in $\SO_3$.  In
this case we claim that the corresponding twisted centraliser is
nothing but $\SU_2^{\alpha_1+\alpha_2}$.  If we denote by
$\{H_{\alpha_1+\alpha_2} , E_{\pm (\alpha_1+\alpha_2)}\}$ the standard
basis of $\su(2)_{\alpha_1+\alpha_2}$, the condition that
$E_{\alpha_1+\alpha_2}$ belong to $T_e\eZ_\tau(h;\SU_3)$, in other
words, that it describe a vector normal to the D-brane is given by
\begin{equation*}
\Ad_{h}E_{\alpha_1+\alpha_2} = \tau(E_{\alpha_1+\alpha_2})~.
\end{equation*}
This can be verified using the fact that $\tau(E_{\alpha_1+\alpha_2})
= -E_{\alpha_1+\alpha_2}$.  A similar argument can be carried out for
the other basis elements of $\su(2)_{\alpha_1+\alpha_2}$.

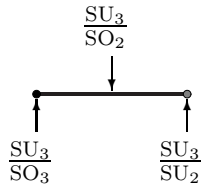
\begin{figure}[h!]
  \begin{center}
    \begin{picture}(40,30)
      \Thicklines
      \put(10.5,15){\line(1,0){19}}
      \thinlines
      \put(30,15){\shade\circle{1}}
      \put(10,15){\circle*{1}}
      \put(20,20){\vector(0,-1){4.5}}
      \put(16,23){$\frac{\SU_3}{\SO_2}$}
      \put(10,10){\vector(0,1){4.3}}
      \put(6,5){$\frac{\SU_3}{\SO_3}$}
      \put(30,10){\vector(0,1){4.3}}
      \put(26,5){$\frac{\SU_3}{\SU_2}$}
    \end{picture}
    \caption{Moduli space of twisted conjugacy classes of $\SU_3$}
    \label{fig:MtcSU3}
  \end{center}
\end{figure}

We thus obtain that the classical moduli space of twisted D-branes in
$\SU_3$ is given by roughly a `one half fraction' of the moduli space
of symmetric D-branes in $\SO_3$.  More precisely,
\begin{equation}\label{eq:clmtcc}
\eM_{cl}(\SU_3,\tau) = \left\{ X\in\gt^\tau \mid 0\leq \bar\alpha(X)\leq
                        \tfrac{1}{4} \right\}~. 
\end{equation}
The point $\bar\alpha(X)=0$ is singular in $\SO_3$ and
$\tau$-singular in $\SU_3$, giving rise to a $5$-dimensional twisted
D-brane of the form $\SU_3/\SO_3$.  The other endpoint
$4\bar\alpha(X)=1$ is regular in $\SO_3$, but $\tau$-singular in
$\SU_3$.  The corresponding twisted class is also $5$-dimensional,
but has the form $\SU_3/\SU_2$.  Finally, the interior points are
regular and give rise to $7$-dimensional twisted conjugacy classes of
the form $\SU_3/\SO_2$.  The resulting space of classical twisted
D-branes in $\SU_3$ is described in Figure~\ref{fig:MtcSU3}.  Notice
that in this case we have that the dimension of these twisted
conjugacy classes is always odd, due to the fact that the difference
between the ranks of $\SU_3$ and $\SO_3$ is odd.

Notice also that in this case, by contrast with the case of standard
conjugacy classes described in Section 2, the action of the centre of
$\SU_3$ on the space of twisted conjugacy classes is trivial since we
have
\begin{equation*}
\eC_\tau(z\cdot X) = \eC_\tau(X)~.
\end{equation*}

\begin{table}[h!]
\centering
\setlength{\extrarowheight}{16pt}
\begin{tabular}{|>{$}c<{$}|>{$}c<{$}|>{$}c<{$}|>{$}c<{$}|}\hline 
\bar\alpha(X)  & \SU_2 & \SO_3 & \SU_3,\tau \\
  & X\quad\quad \eC(X) & X\quad\quad\eC(X)
& X\quad\quad\ \eC_\tau(X)\\ \hline\hline 
0 & \text{sing}~,\quad \frac{\SU_2}{\SU_2} & \text{sing}~,\quad 
\frac{\SO_3}{\SO_3} & \text{$\tau$-sing}~,\quad \frac{\SU_3}{\SO_3}\\
  & \text{reg}~,\quad \frac{\SU_2}{\U(1)} & \text{reg}~,\quad
  \frac{\SO_3}{\SO_2} & \text{$\tau$-reg}~,\quad \frac{\SU_3}{\SO_2} \\
\frac 14 & \text{reg}~,\quad \frac{\SU_2}{\U(1)} & \text{reg}~,\quad
\frac{\SO_3}{\SO_2} & \text{$\tau$-sing}~,\quad \frac{\SU_3}{\SU_2}\\
  & \text{reg}~,\quad \frac{\SU_2}{\U(1)} & \text{reg}~,\quad
  \frac{\SO_3}{\SO_2} & \\
\frac 12 & \text{reg}~,\quad \frac{\SU_2}{\U(1)} & \text{reg}~,\quad
  \frac{\SO_3}{\O_2} & \eC_\tau(X)=\eC_\tau(\frac{\bar\alpha^*}{4}-X)\\
  & \text{reg}~,\quad \frac{\SU_2}{\U(1)} &  & \\
\frac 34 & \text{reg}~,\quad \frac{\SU_2}{\U(1)} &
\eC(X)=\eC(\frac{\bar\alpha^*}{2}-X) & \eC_\tau(X) =
\eC_\tau(X+\frac{\bar\alpha^*}{4})\\ 
 & \text{reg}~,\quad \frac{\SU_2}{\U(1)} &  & \\
1 & \text{sing}~,\quad \frac{\SU_2}{\SU_2} &  & \\
\hline
\end{tabular}
\vspace{8pt}
\caption{Comparison between conjugacy classes in $\SU_2$, $\SO_3$
and twisted conjugacy classes in $\SU_3$. (Here, $\SO_3=\SU_3^\tau$ 
and $\SU_2$ is its double cover.)}
\label{tab:conj-classes}
\end{table}

Let us also mention that in the case of $\SU_3$ one can determine the
twisted conjugacy classes by direct calculation.  If we can think of
$\SU_3$ as the group of special unitary $3\times 3$ matrices, it is
convenient to use a different representative for the outer
automorphism, which is given by complex conjugation, that is, $\rho(g)
= \bar g$, for any $g$ in $\SU_3$.  Clearly, $\tau$ and $\rho$ are
related by an inner automorphism, which is equivalent, at the level of
the group, to a change of coordinates.  The fixed point subgroup in
this case is the $\SO_3$ subgroup of $\SU_3$ given by the real valued
matrices $g$.  Its maximal torus can be parametrised as follows:
\begin{equation}\label{eq:par}
T(\SO_3) = \left\{ h_\theta = \begin{pmatrix}
                               \cos 2\theta & \sin 2\theta & 0\\
                              -\sin 2\theta & \cos 2\theta & 0\\
                                    0       &      0       & 1
                          \end{pmatrix} \Bigg\vert \theta\in[-\frac{\pi}{2} ,
                              \frac{\pi}{2}] \right\}~.
\end{equation}
In particular, one can easily check that
\begin{equation*}
\eC_\rho(h_\theta) = \eC_\rho(h_{\pi/2-\theta}) =
\eC_\rho(h_{\theta+\pi/2}) = \eC_\rho(h_{-\theta})~,
\end{equation*}
which is a manifestation of the action of the twisted Weyl group.
This implies that the space of twisted conjugacy classes is
parametrised by $\theta\in[0,\pi/4]$, with
$\eC_\rho(h_0)\cong\SU_3/\SO_3$, $\eC_\rho(h_\theta)\cong\SU_3/\SO_2$
for $0<\theta<\pi/4$, and $\eC_\rho(h_{\pi/4})\cong\SU_3/\SU_2$.
Also, one can explicitly check in this case that the elements
$h_\theta$, for $\theta=\pi/4$, and $\theta=\pi/2$, despite being
regular in $\SO_3$, possess nontrivial twisted centralisers, and give
rise to lower-dimensional twisted D-branes.

We find it instructive to compare, in Table~\ref{tab:conj-classes},
the moduli spaces of conjugacy classes in $\SU_2$ and $\SO_3$ with the
moduli space of twisted conjugacy classes in $\SU_3$, as all three of
them can be described in terms of the same parameter.  This example
illustrates the fact that the notion of regularity in $G^\tau$ is not
appropriate for describing the space of twisted conjugacy classes;
instead, the relevant notion is that of $\tau$-regularity.  It is also
important to remark that, although twisted conjugacy classes are,
roughly speaking, parametrised by $T^\tau_0$, regular elements in
$T^\tau_0$ do \emph{not} necessarily give rise to top-dimensional
twisted classes of the form $G/T^\tau_0$, as previously stated in the
literature \cite{FFFS}.

\subsection{Quantum analysis}

The classical analysis of the previous paragraph gave us the space
\eqref{eq:clmtcc} of twisted conjugacy classes in $\SU_3$, which forms
a continuous family parametrised by the points of an interval in
$\gt^\tau$.  In order to determine the twisted D-brane configurations
that are consistent at the quantum level, we must analyse the
quantisation conditions imposed by the requirement of
single-valuedness of the path integral.  In this case, the global
worldsheet anomaly is measured by the periods of $(H,\omega)/2\pi$
over the relative cycles in $H_3(\SU_3,\eC_\tau)$, where $H$ is the
same three-form field of the $\SU_3$ WZW model, whereas the two-form
field $\omega$ defined on the D-brane is the one determined in
\cite{Q0} for a general automorphism.

\begin{figure}[h!]
  \setlength{\unitlength}{0.65mm}
  \centering
  \subfigure[]{
    \begin{picture}(60,65)(0,-5)
      \thinlines
      \put(25,0){\vector(0,1){60}}
      \put(27,58){\scriptsize{$\bar\alpha(X)$}}
      \put(21,53){\scriptsize{$\frac 12$}}
      \put(21,-4){\scriptsize{$0$}}
      \thicklines
      \put(0,0){\color{light2}
        \put(25,10){\arc{20}{0}{3.14159}}
        \put(15,10){\line(0,1){40}}
        \put(35,10){\line(0,1){40}}
        \put(15,50){\line(1,0){20}}
        }
      \Thicklines
      \put(0,0){\line(1,0){35}}
      \put(0,50){\line(1,0){35}}
      \put(35,25){\arc{50}{4.71239}{7.85398}}
      \put(2,2){\scriptsize{$\eC_\tau$}}
    \end{picture}
    }
  \goodgap
  \subfigure[]{
    \begin{picture}(50,65)(0,-5)
      \thinlines
      \put(25,0){\vector(0,1){60}}
      \put(27,58){\scriptsize{$\bar\alpha(X)$}}
      \put(21,53){\scriptsize{$\frac 12$}}
      \put(21,-4){\scriptsize{$0$}}
      \thicklines
      \put(0,0){\color{light2}
        \put(25,10){\arc{20}{0}{3.14159}}
        \put(15,10){\line(0,1){40}}
        \put(35,10){\line(0,1){40}}
        \put(15,50){\line(1,0){20}}
        }
      \Thicklines
      \put(0,10){\line(1,0){35}}
      \put(0,40){\line(1,0){35}}
      \put(35,25){\arc{30}{4.71239}{7.85398}}
      \put(2,12){\scriptsize{$\eC_\tau$}}
      \put(27,12){\scriptsize{$\eC$}}
      \put(27,35){\scriptsize{$\eC'$}}
    \end{picture}
    }
  \goodgap
  \subfigure[]{
    \begin{picture}(40,65)(0,-5)
      \thinlines
      \put(25,0){\vector(0,1){60}}
      \put(27,58){\scriptsize{$\bar\alpha(X)$}}
      \put(21,53){\scriptsize{$\frac 12$}}
      \put(21,28){\scriptsize{$\frac 14$}}
      \put(21,-4){\scriptsize{$0$}}
      \thicklines
      \put(0,0){\color{light2}
        \put(25,10){\arc{20}{0}{3.14159}}
        \put(15,10){\line(0,1){40}}
        \put(35,10){\line(0,1){40}}
        \put(15,50){\line(1,0){20}}
        }
      \Thicklines
      \put(0,25){\line(1,0){35}}
      \put(2,27){\scriptsize{$\eC_\tau$}}
    \end{picture}
    }
  \caption{Twisted conjugacy classes in $\SU_3$ and their
    intersections with the fixed point subgroup $\SO_3$:
    (a)~$\eC_\tau \protect\cong \SU_3/\SO_3$, (b)~$\eC_\tau
    \protect\cong \SU_3/\SO_2$, (c)~$\eC_\tau \protect\cong
    \SU_3/\SU_2$.}
  \label{fig:tintersection}
\end{figure}
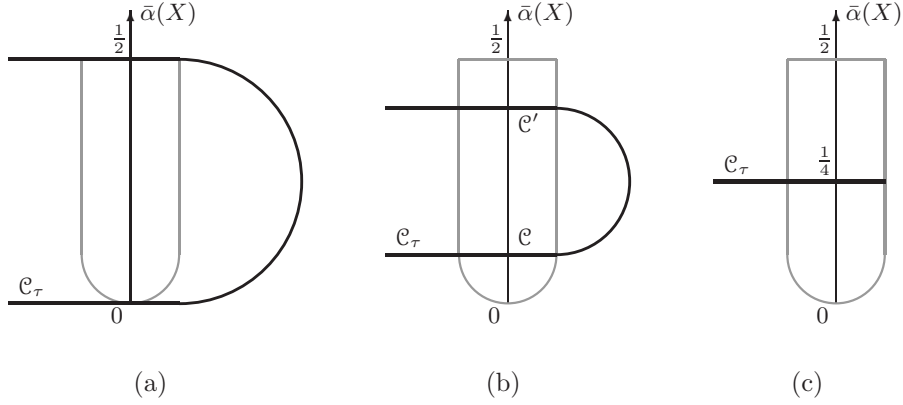

Our strategy for analysing the quantisation conditions in this case
will be to try and make use of the knowledge we have acquired about
the consistent D-branes in $\SO_3$.  More precisely, we will show how
the quantisation conditions satisfied by a D-brane in $\SU_3$
wrapping a twisted conjugacy class $\eC_\tau(h)$, where $h = \exp(X)$,
for some $X$ in $\gt^\tau$, can be evaluated in terms of the
quantisation conditions for D-branes wrapping standard conjugacy
classes in the fixed point subgroup $\SO_3$.  To this end, let us
determine the intersection of the twisted conjugacy class
$\eC_\tau(h)$ with this $\SO_3$.  We first of all notice that the
standard conjugacy class $\eC(h)$ in $\SO_3$, corresponding to the
same element $h$ in the maximal torus of $\SO_3$, is a submanifold of
$\eC_\tau(h)$, since $\tau$ restricts to the identity on $\SO_3$:
\begin{equation*}
\{ \tau(g)h g^{-1} \mid g\in\SO_3 \} = \eC(h;\SO_3)~.
\end{equation*}
Contrary to appearance, this does not fully account for the
intersection we are trying to determine.  In order to see this, we
recall that $X$ and $\alpha^*_T-X$ are related by a twisted Weyl
transformation, which implies that
\begin{equation*}
\eC_\tau(X) = \eC_\tau(\alpha^*_T-X)~.
\end{equation*}
Since, on the other hand, $X$ and $\alpha^*_T-X$ are \emph{not}
related by a Weyl transformation, this indicates that our twisted 
conjugacy class $\eC_\tau$ intersects $\SO_3$ in two generically
distinct conjugacy classes of $\SO_3$:
\begin{equation}\label{eq:tintersection}
\eC_\tau(X;\SU_3) \cap \SO_3 = \eC(X;\SO_3) \cup
                                 \eC(\alpha^*_T-X;\SO_3)~.
\end{equation}
Depending on $X$, we have three cases, which are schematically drawn
in Figure~\ref{fig:tintersection}.  For $X$ such that
$\bar\alpha(X)=0$, the intersection of the $5$-dimensional $\eC_\tau$
with $\SO_3$ consists in the point-like conjugacy class of the
identity $\eC(e)$ and the non-orientable $2$-dimensional conjugacy
class $\eC \cong \SO_3/\O_2$. For generic $X$, the intersection
consists of two spherical conjugacy classes in $\SO_3$, denoted for
simplicity by $\eC$ and $\eC'$ corresponding to $X$ and
$\alpha^*_T-X$, respectively.  Finally, for $\bar\alpha(X)=1/4$,
$\eC_\tau$ intersects $\SO_3$ in only one conjugacy class $\eC \cong
\SO_3/\SO_2$.

The fact that $\eC_\tau$ intersects the fixed point subgroup $\SO_3$
in what turn out to be conjugacy classes of $\SO_3$ prompts us to
investigate the possibility of describing the relative cycles of
$H_3(\SU_3,\eC_\tau)$ in terms of those of $H_3(\SO_3,\eC)$, which we
already know.  Indeed, in the previous section we have seen that for
the generic $\eC \cong \SO_3/\SO_2$, $H_3(\SO_3,\eC) \cong \Z\oplus\Z$
and the generating $3$-cycles are $\SO_3$ itself and the relative
cycle denoted by $(N,\d N)$, where $N$ is given by a $3$-submanifold
in $\SO_3$ such that $\d N=\eC$.  We claim that these two $3$-cycles
can be promoted to relative cycles in $H_3(\SU_3,\eC_\tau) \cong
\Z\oplus\Z$.  In order to show this, we notice that $\d
N=\eC\subset\eC_\tau$; furthermore one has to check that $H$ and
$(H,\omega)$ do not vanish when integrated on $i(\SO_3)$ and $(i(N),\d
i(N))$, respectively, where $i:\SO_3\to\SU_3$ denotes the embedding of
$\SO_3$ in $\SU_3$.  This is indeed true, as we will see in a moment.

The fact that every relative cycle in $H_3(\SO_3,\eC)$ gives rise to
one in $H_3(\SU_3,\eC_\tau)$ implies that a necessary condition for
a twisted brane to wrap $\eC_\tau$ is that the global worldsheet
anomaly evaluated on the `induced' relative cycle should take integer
values.  In other words, the following quantisation conditions must be 
be satisfied:
\begin{equation}\label{tqc1}
\frac{1}{2\pi}\int_{i(\SO_3)} H\ \in\ \Z~,
\end{equation}
for the honest $3$-cycle, and 
\begin{equation}\label{eq:tqc2}
\frac{1}{2\pi}\left(\int_{i(N)} H - \int_{\d i(N)} \omega\right)\ \in\
                \Z~,
\end{equation}
for the second relative cycle.  A brief inspection of these conditions
shows that they are nothing but the quantisation conditions for the
admissible brane configurations in the fixed point subgroup WZW
theory, that is, in $\SO_3$, whose fields are given by $H_{\SO_3}=i^*
H$ and $\omega_{\SO_3} = i^*\omega$.

In order to evaluate the first quantisation condition we use the fact
that the `induced' $\SO_3$ WZW model on the fixed point group has a
level $k_{\SO_3}$ given by\footnote{Alternatively, as done in
  Appendix~\ref{sec:homology}, one can show that $\SO_3$ is twice the
  generator of $H_3(\SU_3)$, which leads us to a similar conclusion
  regarding the level of the $\SO_3$ theory.}
\begin{equation*}
k_{\SO_3} = k \frac{(\alpha_3,\alpha_3)}{(\bar\alpha,\bar\alpha)} =
             4k~,
\end{equation*}
where, we recall, $\alpha_3=\alpha_1+\alpha_2$ is the maximal root of
$\SU_3$.  We thus deduce that 
\begin{equation*}
\frac{1}{2\pi}\int_{i(\SO_3)} H = \frac{1}{2\pi}\int_{\SO_3}
                                   H_{\SO_3} = 2k~,
\end{equation*}
where $H_{\SO_3} = i^* H$ is the three-from field of the `induced' WZW
model on $\SO_3$.  In light of this result, the quantisation condition
corresponding to the honest $3$-cycle reiterates the condition that
the level be an integer, $k\in \Z$.  The second quantisation condition
can be evaluated using a similar line of argument and the calculation
of the global worldsheet anomaly for the $\SO_3$ theory performed in
the previous section, thus obtaining
\begin{equation*}
\frac{1}{2\pi}\left(\int_{i(N)} H - \int_{\d i(N)} \omega\right) = 
\frac{1}{2\pi}\left(\int_N H_{\SO_3} - \int_{\d N}
\omega_{\SO_3}\right) = 4k\bar\alpha(X)~.
\end{equation*}
Hence the consistent D-branes wrapping twisted conjugacy classes in
$\SU_3$ are characterised by integral values of $4k\bar\alpha(X)$.

Knowing that the relative cycles of the fixed point subgroup can be
promoted to relative cycles in $\SU_3$ is however not enough if we
want to accurately describe the quantisation conditions for the
twisted branes.  What we need is to be able to exhibit a basis of
generators for the relative homology group $H_3(\SU_3,\eC_\tau)$, for
a fixed $\eC_\tau$, and evaluate the global worldsheet anomaly on
these generators.  It turns out that the most convenient way of
solving this problem is to resort to a different representative for
the outer automorphism of $\SU_3$, namely the one provided by complex
conjugation $\rho$.  This is done in Appendix~B, where the generators
of $H_3(\SU_3,\eC_\rho)$ are determined and the corresponding
quantisation conditions are derived.  Using the fact that $\tau$ and
$\rho$ are related by an inner automorphism, we deduce that the
correct quantisation condition for $\eC_\tau(X)$ reads
\begin{equation}\label{eq:tqc}
2k\bar\alpha(X) = 2m - k~,\qquad\qquad m\in\Z~.
\end{equation}

Let us now consider separately the quantisation conditions
corresponding to the two `singular' twisted D-branes.  Let us begin
with the twisted class of the identity, $\eC_\tau\cong
\SU_3/\SO_3$.  The relevant relative homology group in this case is
$H_3(\SU_3,\eC_\tau)\cong\Z$, where the $\SO_3$ subgroup is four
times the generator.  We thus obtain the following quantisation
condition
\begin{equation*}
\frac{1}{2\pi}\int_{i(\SO_3)} H\ \in\ 4\Z~,
\end{equation*}
which imposes that $k\in 2\Z$.  In other words, this particular state
only appears as a consistent configuration for even values of the
level.  Finally, for the other $5$-dimensional twisted brane wrapping
$\eC_\tau \cong \SU_3/\SU_2$, we have $H_3(\SU_3,\eC_\tau)\cong\Z$,
with the $\SO_3$ subgroup being twice the generator, so that the
quantisation condition reads
\begin{equation*}
\frac{1}{2\pi}\int_{i(\SO_3)} H\ \in\ 2\Z~,
\end{equation*}
imposing that the level $k$ be an integer.

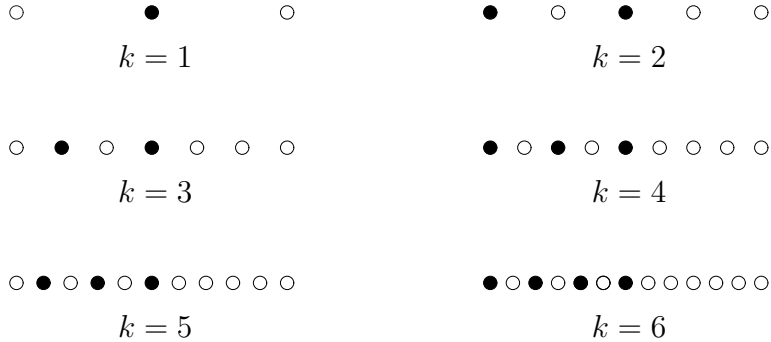
\begin{figure}[h!]
  \setlength{\unitlength}{0.9mm}
  \begin{center}
    \begin{picture}(120,60)
      \put(5,50){%
        \put(0,0){\circle{2}}
        \put(20,0){\circle*{2}}
        \put(40,0){\circle{2}}
        \put(15,-8){\makebox{$k=1$}}
        }
      \put(75,50){%
        \put(0,0){\circle*{2}}
        \put(10,0){\circle{2}}
        \put(20,0){\circle*{2}}
        \put(30,0){\circle{2}}
        \put(40,0){\circle{2}}
        \put(15,-8){\makebox{$k=2$}}
        }
      \put(5,30){%
        \put(0,0){\circle{2}}
        \put(6.6667,0){\circle*{2}}
        \put(13.3333,0){\circle{2}}
        \put(20,0){\circle*{2}}
        \put(26.6667,0){\circle{2}}
        \put(33.3333,0){\circle{2}}
        \put(40,0){\circle{2}}
        \put(15,-8){\makebox{$k=3$}}
        }
      \put(75,30){%
        \put(0,0){\circle*{2}}
        \put(5,0){\circle{2}}
        \put(10,0){\circle*{2}}
        \put(15,0){\circle{2}}
        \put(20,0){\circle*{2}}
        \put(25,0){\circle{2}}
        \put(30,0){\circle{2}}
        \put(35,0){\circle{2}}
        \put(40,0){\circle{2}}
        \put(15,-8){\makebox{$k=4$}}
        }
      \put(5,10){%
        \put(0,0){\circle{2}}
        \put(4,0){\circle*{2}}
        \put(8,0){\circle{2}}
        \put(12,0){\circle*{2}}
        \put(16,0){\circle{2}}
        \put(20,0){\circle*{2}}
        \put(24,0){\circle{2}}
        \put(28,0){\circle{2}}
        \put(32,0){\circle{2}}
        \put(36,0){\circle{2}}
        \put(40,0){\circle{2}}
        \put(15,-8){\makebox{$k=5$}}
        }
      \put(75,10){%
        \put(0,0){\circle*{2}}
        \put(3.3333,0){\circle{2}}
        \put(6.6667,0){\circle*{2}}
        \put(10,0){\circle{2}}
        \put(13.3333,0){\circle*{2}}
        \put(16.6667,0){\circle{2}}
        \put(20,0){\circle*{2}}
        \put(13.3333,0){\circle{2}}
        \put(16.6667,0){\circle{2}}
        \put(23.3333,0){\circle{2}}
        \put(26.6667,0){\circle{2}}
        \put(30,0){\circle{2}}
        \put(33.3333,0){\circle{2}}
        \put(36.6667,0){\circle{2}}
        \put(40,0){\circle{2}}
        \put(15,-8){\makebox{$k=6$}}
        }
    \end{picture}
    \caption{Quantum moduli space for $\tau$-twisted D-branes in
      $\SU_3$ at level $k$ (black circles) compared with that of
      $\SU_3^\tau \protect\cong \SO_3$ (white circles) at level $4k$.}
    \label{fig:MqSU3SO3}
  \end{center}
\end{figure}

We thus obtain that the space of twisted D-branes in $\SU_3$ is
given by
\begin{equation}\label{eq:qmstau}
\eM_q(\SU_3,\tau) = \begin{cases}
\{X\in\gt^\tau \mid 4k\bar\alpha(X) = 1,3,...,k\}~, &\text{for}\quad
                                      k\ \text{odd}~,\\ 
\{X\in\gt^\tau \mid 4k\bar\alpha(X) = 0,2,...,k\}~, &\text{for}\quad
                                      k\ \text{even}~,
                    \end{cases}
\end{equation}
and the states corresponding to the first few values of the level are
represented in Figure~\ref{fig:MqSU3SO3}.  At a given odd level $k$ we
have $\half(k-1)$ $7$-dimensional and one $5$-dimensional branes,
whereas for an even level $k$ we have $(\half k - 1)$ $7$-dimensional
and two $5$-dimensional branes.

If we now compare our quantisation condition for the $7$-dimensional
branes \eqref{eq:tqc} and the additional conditions obtained for the
$5$-dimensional ones with the spectrum \eqref{eq:ihw} of IHW
representations of the twisted affine Lie algebra
$\widehat\su(3)^{(2)}_k$ we can conclude that the admissible twisted
D-brane configurations in $\SU_3$ are in one-to-one correspondence
with the IHW representations of the corresponding twisted affine Lie
algebra $\widehat\su(3)^{(2)}_k$.

In particular, this implies that for even $k$ the twisted D-branes
intersect the fixed point subgroup in $\SO_3$ D-brane configurations
corresponding to even highest weights, whereas for $k$ odd the
corresponding $\SO_3$ branes correspond to odd highest weights.

\section{D-brane charges}

In the previous sections we have analysed in a systematic fashion
which are the possible D-brane configurations in $\SU_3$ described by
the gluing conditions \eqref{eq:gc}.  The next important question to
be addressed is: How can one classify these D-brane configurations?

One possible classification is provided by the very result of our
previous analysis.  Namely, we have seen that the consistent (twisted)
D-brane configurations are in one-to-one correspondence with the
integrable representations of the corresponding (twisted) affine Lie
algebra.  This can be rephrased be saying that every such brane 
configuration is characterised by a set of $\rank(G^r)$ quantum
numbers, where $r$ stands for an arbitrary group automorphism.  These
quantum numbers are obtained as a result of imposing that the path
integral be well defined; they are the possible values taken by the
quantity 
\begin{equation}\label{eq:topch}
\frac{1}{2\pi}\left(\int_N H - \int_{\d N} \omega\right)~,
\end{equation}
where $(N,\d N)$ are taken to be the generators of the appropriate
relative homology group.  Thus \eqref{eq:topch} can be thought of as
defining a \emph{topological} charge whose values measure the quantum
numbers characterising the state of the system.  In the framework of
BCFT, they are the quantum numbers which describe the possible
boundary conditions.  Notice also that \eqref{eq:topch} can be
understood both algebraically and geometrically.  From an algebraic
point of view, we have seen that this describes the highest weights of
the integrable representations associated to a given quantum
configuration.  On the other hand, from a geometric point of view, the
vanishing of the global worldsheet anomaly translates into a set of
quantisation conditions imposed on the \emph{boundary} values of the
fields normal to the D-brane.  This is a typical feature of
topological charges.

In Section 2 we have seen that the space of allowed D-brane
configurations falls into triplets, where the states belonging to one
such triplet describe D-branes that wrap shifted conjugacy classes.
Since shifted conjugacy classes are indistinguishable from one
another, in their intrinsic properties, it is natural to demand that
the charges carried by them be equal.  This suggests that, in order to
obtain a suitable notion of charge, a further reduction prescription
is necessary.

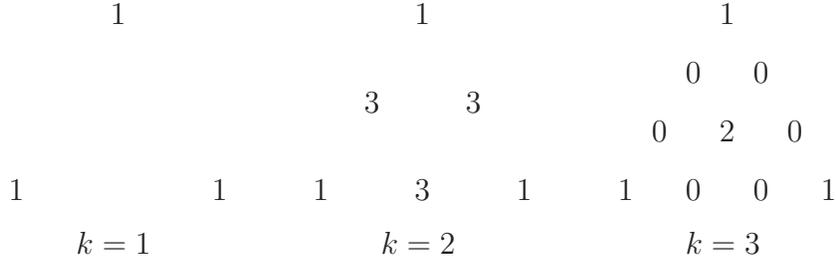
\begin{figure}[h!]
  \setlength{\unitlength}{0.9mm}
  \begin{center}
    \begin{picture}(130,40)
      \put(5,10){%
        \put(0,0){1}
        \put(30,0){1}
        \put(15,25.9808){1}
        \put(10,-8){\makebox{$k=1$}}
        }
      \put(50,10){%
        \put(0,0){1}
        \put(15, 0){3}
        \put(7.5, 12.9904){3}
        \put(22.5, 12.9904){3}
        \put(30,0){1}
        \put(15,25.9808){1}
        \put(10,-8){\makebox{$k=2$}}
        }
      \put(95,10){%
        \put(0,0){1}
        \put(10,0){0}
        \put(20,0){0}
        \put(30,0){1}
        \put(5, 8.66025){0}
        \put(10, 17.3205){0}
        \put(15, 8.66025){2}
        \put(20, 17.3205){0}
        \put(25, 8.66025){0}
        \put(15,25.9808){1}
        \put(10,-8){\makebox{$k=3$}}
        }
    \end{picture}
    \caption{D-brane charges for symmetric branes in $\SU_3$ for
    lowest values of the level $k$.}
    \label{fig:QsSU3}    
  \end{center}
\end{figure}

In the case of untwisted D-branes, where the admissible configurations
are in one-to-one correspondence with the IHW representations of the
affine Lie algebra, it has been argued from various points of view
(see, for instance, \cite{FredSch,MMS2}) that it is natural to define
a charge which is given by the dimension of the IHW representation of
the horizontal subalgebra.  Thus in the case of $\SU_3$, a D-brane
configuration described by a $\widehat\su(3)^{(1)}_k$ IHW
representation is labelled by the $\su(3)$ IHW representation
$(\lambda_1,\lambda_2)$ whose dimension is given by
\begin{equation*}
\dim(\lambda_1,\lambda_2) = \half (\lambda_1 + 1)(\lambda_2 + 1)
(\lambda_1 + \lambda_2 + 2)~.
\end{equation*}
If one now takes into account the equivalence relation
\eqref{eq:equiv} described in Section 2, which translates into the
physical requirement that branes corresponding to shifted conjugacy
classes should carry the same charge, one obtains \cite{MMS2} that the
charge $Q_{(\lambda_1,\lambda_2)}$ of this particular brane is given
by
\begin{equation*}
Q_{(\lambda_1,\lambda_2)} = \dim(\lambda_1,\lambda_2)\mod n(k)~,
\end{equation*}
where
\begin{equation}
n(k) = \begin{cases}
          \ k+3~, &\text{for}\quad k\ \text{even}~,\\ 
          \ \half(k+3)~, &\text{for}\quad k\ \text{odd}~. 
                    \end{cases}
\end{equation}
This completely determines the charges of the $\SU_3$ branes wrapping
conjugacy classes, which turn out to fit in the twisted K-theory group 
\cite{FredSch,MMS2}
\begin{equation*}
K^0_H(\SU_3) = \begin{cases}
                  \Z/(k+3)\Z~, &\text{for}\quad k\ \text{even}~,\\ 
                  \Z/\half(k+3)\Z~, &\text{for}\quad k\ \text{odd}~.
               \end{cases}
\end{equation*}
Furthermore, one can easily see that the admissible branes at a given
level $k$ fall into $k$ multiplets characterised by equal values of
the charge.  Generically, these multiplets consist of $6$ different
states, although there exist multiplets of $1$ or $3$ states as well.
The distribution of the various charges for the first few values of
the level is represented in Figure~\ref{fig:QsSU3}.

Let us now analyse the case twisted case.  In the previous section we
have shown that a twisted brane configuration can be uniquely
characterised by an IHW representation $\hat\mu$ of the twisted affine
Lie algebra $\widehat\su(3)^{(2)}_k$ which, in turn, is labelled by a
IHW representation $\mu$ of the fixed point subalgebra $\so(3)$.
Moreover, since the action of the centre of $\SU_3$ on the space of
twisted conjugacy classes is trivial, in this case we do not have
multiplets of shifted twisted conjugacy classes.  This prompts us to
define the charge of such a D-brane to be given by the dimension of
the corresponding representation of $\so(3)$
\begin{equation*}
Q_{\mu} = \dim(\mu)~.
\end{equation*}
Notice that for these configurations no further reduction is
necessary.  It is easy to explicitly compute the charges of the
allowed D-branes at level $k$ and obtain:
\begin{equation}
Q_\mu = \begin{cases}
             1,3,...,k+1~, &\text{for}\quad k\ \text{even}~,\\
             2,4,...,k+1~, &\text{for}\quad k\ \text{odd}~.
        \end{cases}
\end{equation}
We now see that these charges fit into the charge group predicted by
twisted K-theory \cite{MMS2}
\begin{equation*} 
K^1_H(\SU_3) = \begin{cases}
               \Z/(k+3)\Z~, &\text{for}\quad k\ \text{even}~,\\ 
               2\Z/(k+3)\Z~, &\text{for}\quad k\ \text{odd}~.
               \end{cases}
\end{equation*}

Let us end this section with a few remarks regarding the D-brane
configurations wrapping the (twisted) conjugacy class of the identity.
It is interesting to notice a certain similarity between these
configurations.  First of all, notice that, both in the twisted and in
the untwisted case, the brane wrapping the conjugacy class of the
identity is the configuration carrying the unit of charge.  However,
in the twisted case, this configuration only exists for even values of
the level.  Moreover, one can show that the two-form field $\omega$,
given by \cite{Q0}
\begin{equation*}
  \omega = -\half~\langle g^{-1}dg\ , \frac{\1 + \Ad_{g^{-1}}\tau}{\1 -
    \Ad_{g^{-1}}\tau}\ g^{-1}dg\rangle~,
\end{equation*}
vanishes on the twisted conjugacy class of the identity.  In order to
see this is suffices to parametrise the point $g$ on $\eC_\tau(e)$
by $g = \tau(h) h^{-1}$, in which case $\omega$ can be written as
\begin{equation*}
\omega = - \la h^{-1}dh,\tau(h^{-1}dh)\ra~,
\end{equation*}
from which one deduces that $\omega$ vanishes on $\eC_\tau(e)$ as it
does, trivially, on $\eC(e)$.  From this it follows, in particular,
that also the three-form field $H$ vanishes, when pulled back to
$\eC_\tau(e)$.

\section{Conclusions and outlook}

In this paper we have constructed a detailed and systematic picture of
the consistent D-branes in $\SU_3$ described by the gluing conditions
\eqref{eq:gc}.  The aim of this paper is twofold.  On the one hand, we
provide a detailed description of the possible branes in $\SU_3$,
which is an important case in its own right, as it can be used as a
building block for type II string backgrounds.  On the other hand, we
introduce a systematic framework for the study of the general case,
which makes the subject of a companion paper \cite{SD2notes}.  The
discussion of the $\SU_3$ case introduces all the ingredients
necessary for the discussion of D-branes in general group manifolds;
moreover, many of the arguments presented here are valid for any
compact, connected, simply connected Lie group.  In fact, it is worth
noting that, in many ways, the $\SU_3$ case is the most intricate
among the simply connected groups, essentially due to the subtleties
related to the outer automorphisms of the $A_{2s}$ groups.

At the classical level, the space of (twisted) D-branes is given by
the fundamental domain of the (twisted) Weyl group.  For standard
conjugacy classes this has a natural description in terms of the
Stiefel diagram.  We have shown that the space of twisted conjugacy
classes can be described in a similar way by constructing a natural
generalisation of the Stiefel diagram which we call `twisted Stiefel
diagram'.  The key notion that allows us to distinguish between the
various twisted conjugacy classes is that of `twisted regularity' ---
$\tau$-regularity, for short --- which, as the name suggests, is a
natural generalisation of the familiar notion of regularity obtained
when the adjoint action is replaced with the twisted adjoint action.
Thus the twisted Stiefel diagram is a figure in the Cartan subalgebra
$\gt^\tau$ of the fixed point subalgebra $\gg^\tau$ which is
constructed as the inverse image, under the exponential map, of the
$\tau$-singular points in the fixed point subset $T^\tau$ of the
maximal torus.

The quantum analysis of the consistent D-brane configurations centers
around the requirement that the path integral of the boundary WZW
model be well defined or, in other words, that we have a vanishing
global worldsheet anomaly.  As we know, this amounts to the condition
that $1/2\pi[(H,\omega)]$ defines a class in $H^3(G,\eC_{(\tau)};\Z)$.
For untwisted D-branes it was known that the quantisation conditions
select a discrete subset of configurations labelled by the IHW
representations of $\hat\gg^{(1)}_k$.  One of the main results here is
that twisted D-branes are uniquely characterised by IHW
representations of the corresponding \emph{twisted} affine Lie
algebra, in our case, $\widehat\su(3)^{(2)}_k$, which offers an
a posteriori justification for the notion of twisted conjugacy class 
introduced in \cite{SDnotes}.

One of the most important consequences of this result is a definition
of the charge of the twisted D-branes, which is given by the dimension
of the representation of the fixed point subalgebra.  An explicit
evaluation of the charges for the set of consistent configurations
determined in Section 4 shows that they fit in the twisted K-theory
group $K^1_H(\SU_3)$.

The approach presented here is ideally suited for determining the
admissible D-submanifolds; that is, the submanifolds on which D-branes
can wrap.  One of the most important challenges for the future is to
construct a suitable generalisation able to discern configurations of
more than one D-brane wrapping a certain (twisted) conjugacy class.
This limitation notwithstanding, our approach shows that the correct
charges for configurations consisting of one D-brane on a (twisted)
conjugacy class can be obtained solely on geometrical (and
topological) grounds, without resorting to any dynamical information.

\section*{Acknowledgements}

Some of the results obtained in this paper were announced in the
conferences \emph{Modern trends in string theory}, Lisbon (July 2001),
\emph{The quantum structure of spacetime and the geometric nature of
  fundamental interactions}, Corfu (September 2001), and
\emph{Mathematical aspects of string theory}, Vienna (November 2001),
and I would like to take this opportunity to thank the organisers.  It
is a pleasure to thank C~Bachas, V~Schomerus and R~Wendt for
stimulating discussions and correspondence, and JM~Figueroa-O'Farrill
for many useful discussions and help with the pictures.  This work was
completed during a visit to the Erwin Schr\"odinger Institute, Vienna,
whom I would like to thank for support.

\appendix

\section{Basic facts about $\widehat\su(3)^{(2)}_k$}
\label{sec:su32k}

We collect here a few known facts about the twisted affine Lie algebra 
$\widehat\su(3)^{(2)}_k$.  For more details see \cite{Kac,GO}.  Let
$\gg$ be the Lie algebra $\su_3$ and $\tau$ the Dynkin diagram
automorphism of $\gg$, such that $\tau^2=\1$.  The finite dimensional
Lie algebra $\gg$ can be split into eigenspaces $\gg_{(m)}$ of
$\tau$ 
\begin{equation*}
\gg = \gg_{(0)} \oplus \gg_{(1)}~,
\end{equation*}
with $\tau|_{\gg_{(0)}} = \1$, $\tau|_{\gg_{(1)}} = -\1$.  Moreover,
we have $[\gg_{(m)},\gg_{(n)}] = \gg_{(m+n)}$.  We know from Section 3
that the fixed point subalgebra $\gg_{(0)}\cong\so_3$, and its simple
root is given by $\bar\alpha = \half(\alpha_1+\alpha_2)$.

The simple roots of the twisted affine Lie algebra
$\widehat\su(3)^{(2)}_k$ are given by 
\begin{align*}
a_{(0)} &= (~ -\phi~,~0~,~\half~)~,\\
a_{(1)} &= (~ \bar\alpha~,~0~,~0~)~,
\end{align*}
where $\phi = \alpha_1+\alpha_2$ is the highest root of $\gg$ and
$\bar\alpha = \half(\alpha_1+\alpha_2)$ is the simple root of
$\gg_{(0)}$.  An IHW representation $\hat\mu$ of
$\widehat\su(3)^{(2)}_k$ is labelled by an IHW representation $\mu$ of
$\gg_{(0)} \cong \so_3$ together with a value of the level $k$, which
we write as $\hat\mu = (\mu,k,0)$.  The conditions for having an IHW
representation for the twisted affine algebra require that the
quantity
\begin{equation*}
\frac{2(a,\mu)}{(a,a)}
\end{equation*}
take positive integral values for any root $a$, and also that
\begin{equation*}
0 \leq 2(\phi,\mu) \leq k~.
\end{equation*}

One thus obtains that the highest weights of $\widehat\su(3)^{(2)}_k$ are 
of the form
\begin{equation*}
n_0 l_{(0)} + n_1 l_{(1)}~,
\end{equation*}
where $l_{(0)}$ and $l_{(1)}$ are the fundamental weights of
$\widehat\su(3)^{(2)}_k$, which read
\begin{align*}
l_{(0)} &= (~0~,~2~,~0~)~,\\
l_{(1)} &= (~\lambda~,~1~,~0~)~,
\end{align*}
with $\bar\lambda = \frac{\bar\alpha}{2}$ the fundamental weight of
$\so_3$.  The positive integers $n_0$ and $n_1$ are determined as
solutions of the equation
\begin{equation*}
\frac{2k}{(\bar\alpha,\bar\alpha)} = M (2 n_0 + n_1)~,
\end{equation*}
where $M=\frac{(\phi,\phi_)}{(\bar\alpha,\bar\alpha)}=4$.  This
equation can be rewritten in the simpler form
\begin{equation}\label{eq:ihw}
k = 2 n_0 + n_1~.
\end{equation}
Thus the IHW representations of $\widehat\su(3)^{(2)}_k$
are labelled by the $\so_3$ highest weights $\mu = n_1\bar\lambda$,
where the integer $n_1$ takes the following values: 
\begin{equation}
n_1 = \begin{cases}
          \ 1,3,...,k~, &\text{for}\quad k\ \text{odd}~,\\ 
          \ 0,2,...,k~, &\text{for}\quad k\ \text{even}~. 
                    \end{cases}
\end{equation}

\section{On the relative (co)homology of $\SU_3$}
\label{sec:homology}
\vspace{-5pt}
\begin{center}
  (with José Figueroa-O'Farrill)
\end{center}
\vspace{10pt}

In this appendix we collect some topological results which are used in
the main body of the paper concerning the (co)homology of $\SU_3$
relative its twisted conjugacy classes.  Let $G=\SU_3$ throughout this
appendix, identified with the group of $3\times 3$ unitary matrices
with unit determinant.  Complex conjugation defines an automorphism
$\rho$ which is not inner and whose fixed point set $G^\rho$ is the
subgroup consisting of $3\times 3$ real orthogonal matrices with unit
determinant; that is, $\SO_3$.  This is a closed submanifold of $G$
and hence a (geometric) 3-cycle defining a homology class in $H_3(G)$.
In the first part of this appendix we will show (by wholly elementary
means) that this is class is twice the generator of $H_3(G)\cong\ZZ$.
In the second part of this appendix we exhibit generators for
$H_3(G,\eC_\rho)$, where $\eC_\rho$ is a twisted conjugacy class and
we also prove that $H_2(G,\eC_\rho)$ vanishes.  In the third and final
part we determine which twisted conjugacy classes are admissible in
the sense of \cite{FSrc}.

JMF would like to acknowledge useful discussions with Andrew Ranicki
and Elmer Rees, the support of the Erwin Schr\"odinger Institute during
the time this work was done, and a travel grant from the PPARC.

\subsection{The homology class of $\SO_3$ in $\SU_3$}

Since $G$ is compact and simple, $H_3(G) \cong \ZZ$ and the
isomorphism is realised by integrating a suitably normalised $3$-form
$\Omega$ on the cycle: $[Z] \mapsto \int_Z \Omega$.  Up to
normalisation, there is a unique harmonic $3$-form on $G$: $\Omega =
\lambda \Tr \left(g^{-1} dg\right)^3$, for some $\lambda$.  Our
strategy is the following:
\begin{itemize}
\item we fix $\lambda$ by demanding that $\int_Z \Omega = 1$, where
  $Z$ is the $\SU_2$ subgroup associated to any one of the roots,
  since any of these subgroups is a generator of $H_3(G)$; and
\item we integrate the suitably normalised $\Omega$ on the $\SO_3$
  subgroup.
\end{itemize}
In summary we will show that the homology class of $\SO_3$ in $G$ is
twice that of the generator of $H_3(G)$.

Consider the $\SU_2$ subgroup of $G$ consisting of matrices of the
form
\begin{equation}
  \label{eq:SU2alpha}
  \begin{pmatrix}
    \phantom{-}a & b & 0\\ -\bar b & \bar a & 0 \\ \phantom{-}0 & 0 & 1
  \end{pmatrix}
\end{equation}
where $|a|^2 + |b|^2 = 1$.  It is a geometric $3$-cycle which
generates $H_3(G)$.  We will parametrise this submanifold of $G$ with
coordinates $(\theta,\phi,\psi)$ as follows
\begin{equation}
  \label{eq:params}
  a = \cos\theta e^{i\phi} \qquad \text{and}\qquad b = \sin\theta
  e^{i\psi}~,
\end{equation}
where $\phi$ and $\psi$ are angles (ranging from $0$ to $2\pi$) and
$\theta$ ranges from $0$ to $\pi/2$, to render $\sin\theta$ and
$\cos\theta$ non-negative, as they play the role of radial
coordinates.

Restricting the left-invariant (for definiteness) Maurer--Cartan
$1$-forms to this subgroup, we find
\begin{equation*}
  g^{-1}dg = \sum_{i=1}^3 \theta^i \tau_i
\end{equation*}
where $\tau_i$ are a basis for the generators of $\su_2$ inside
$\su_3$.  Explicitly, we have
\begin{equation}
  \label{eq:taus}
  \tau_1 = 
  \begin{pmatrix}
    0 & -i & 0\\ -i & 0 & 0 \\ 0 & 0 & 0
  \end{pmatrix}\qquad
  \tau_2 = 
  \begin{pmatrix}
    0 & -1 & 0 \\ 1 & 0 & 0 \\ 0 & 0 & 0
  \end{pmatrix}\qquad
  \tau_3 = 
  \begin{pmatrix}
    -i & 0 & 0\\ 0 & i & 0 \\ 0 & 0 & 0
  \end{pmatrix}~,
\end{equation}
and a calculation reveals that
\begin{equation}
  \label{eq:thetas}
  \begin{aligned}[m]
    \theta^1 &= \sin(\phi-\psi) d\theta - \half \sin(2\theta)
    \cos(\phi-\psi) (d\phi + d\psi)\\
    \theta^2 &= -\cos(\phi-\psi) d\theta - \half \sin(2\theta)
    \sin(\phi-\psi) (d\phi + d\psi)\\
    \theta^3 &= - (\cos\theta)^2 d\phi + (\sin\theta)^2 d\psi~,
  \end{aligned}
\end{equation}
from where it follows that
\begin{equation*}
  \Omega = \lambda \Tr \left(g^{-1} dg\right)^3 = 6\lambda
  \sin(2\theta) d\theta \wedge d\phi \wedge d\psi~.
\end{equation*}
Integrating this over the range of our coordinates, we find
\begin{equation*}
  \int_{\SU_2} \Omega = \lambda 24 \pi^2~,
\end{equation*}
whence
\begin{equation}
  \label{eq:genH3}
  \Omega = \frac{1}{24\pi^2} \Tr \left(g^{-1} dg\right)^3
\end{equation}
is a de~Rham representative for the generator of $H^3(G)$.

We will now parametrise the $\SO_3$ subgroup consisting of real
matrices in $G$.  It is convenient to realise $\SO_3$ as the adjoint
group of $\SU_2$ acting on the Lie algebra $\su_2$.  In other
words, let $\Ad:\SU_2 \to \SO_3$ be the adjoint action.  Relative to
the basis $\tau_i$ for $\su_2$, we can write the entries of the
adjoint matrix $\Ad(g)$ of an element $g$ in $\SU_2$ as
\begin{equation*}
  \Ad(g)_{ij} = -\half \Tr \left(g \tau_j g^{-1} \tau_i\right)~.
\end{equation*}
For
\begin{equation*}
  g = 
  \begin{pmatrix}
    \phantom{-}\cos\theta e^{i\phi} & \sin\theta e^{i\psi} \\
    -\sin\theta e^{-i\psi} & \cos\theta e^{-i\phi}
  \end{pmatrix}
\end{equation*}
the adjoint matrix $\Ad(g)$ is
\begin{scriptsize}
  \begin{equation*}
    \begin{pmatrix}
      (\cos\theta)^2 \cos(2\phi) - (\sin\theta)^2 \cos(2\psi) &
      (\cos\theta)^2 \sin(2\phi) + (\sin\theta)^2 \sin(2\psi) &
      -\sin(2\theta)\cos(\phi+\psi)\\
      (\sin\theta)^2 \sin(2\psi) - (\cos\theta)^2 \sin(2\phi) &
      (\cos\theta)^2 \cos(2\phi) + (\sin\theta)^2 \cos(2\psi) &
      \sin(2\theta) \sin(\phi+\psi)\\
      \cos(\phi-\psi)\sin(2\theta) & \sin(\phi-\psi)\sin(2\theta) &
      \cos(2\theta)
    \end{pmatrix}~.
  \end{equation*}
\end{scriptsize}
Because $\Ad$ is a two-to-one map, this parametrisation covers
$\SO_3$ twice.

Pulling back the Maurer--Cartan form to $\SU_2$, we have
\begin{equation*}
  \Ad(g)^{-1} d\Ad(g) = \sum_{i=1}^3 \Ad^*\theta^i \Ad_*\tau_i~,
\end{equation*}
where
\begin{align*}
  \Ad^*\theta^1 &= \sin(\psi-\phi) d\theta + \half \sin(2\theta)
  \cos(\psi-\phi) (d\phi + d\psi)\\
  \Ad^*\theta^2 &= \cos(\psi-\phi) d\theta + \half \sin(2\theta)
  \sin(\psi-\phi) (d\phi + d\psi)\\
  \Ad^*\theta^3 &= (\cos\theta)^2 d\phi - (\sin\theta)^2 d\psi~,
\end{align*}
and
\begin{equation*}
  \Ad_* \tau_1 = 
  \begin{pmatrix}
    0 & 0 & 0\\ 0 & 0 & 2 \\ 0 & -2 & 0
  \end{pmatrix}\qquad
  \Ad_* \tau_2 = 
  \begin{pmatrix}
    0 & 0 & -2 \\ 0 & 0 & 0 \\ 2 & 0 & 0
  \end{pmatrix}\qquad
  \Ad_* \tau_3 = 
  \begin{pmatrix}
    0 & 2 & 0\\ -2 & 0 & 0 \\ 0 & 0 & 0
  \end{pmatrix}~.
\end{equation*}
Therefore the pull-back of the normalised $3$-form $\Omega$ is
\begin{equation*}
  \Ad^* \Omega = \frac{1}{\pi^2} \sin(2\theta) d\theta \wedge d\phi
  \wedge d\psi~.
\end{equation*}
Integrating this over $\SU_2$ we find
\begin{equation*}
  \int_{\SU_2} \Ad^* \Omega = 4~.
\end{equation*}
On the other hand,
\begin{equation*}
  \int_{\SU_2} \Ad^* \Omega = \int_{\Ad(\SU_2)} \Omega = 2
  \int_{\SO_3} \Omega~,
\end{equation*}
whence we conclude that
\begin{equation*}
  \int_{\SO_3} \Omega = 2~.
\end{equation*}

\subsection{On $H_2(\SU_3,\eC_\rho)$ and $H_3(\SU_3,\eC_\rho)$}

In this section we exhibit generators for the relative homology group
$H_3(G,\eC_\rho)$, where $i:\eC_\rho\into G$ is a twisted conjugacy class.  We also
show that $H_2(G,\eC_\rho)$ vanishes.  As shown in the body of the paper,
twisted conjugacy classes are parametrised by (a quotient of) the
maximal torus of the fixed point subgroup $G^\rho = \SO_3$.  More
concretely, let $\eC_\rho(\vartheta)$, for $0\leq \vartheta \leq \pi/4$,
denote the twisted conjugacy class of the element
\begin{equation}
  \label{eq:xtheta}
  h_\vartheta:=
  \begin{pmatrix}
    \phantom{-} \cos 2\vartheta & \sin 2\vartheta & 0 \\
    -\sin 2\vartheta & \cos 2\vartheta & 0 \\
    0 & 0 & 1
  \end{pmatrix}~.
\end{equation}
Then these exhaust all the twisted conjugacy classes of $G$.  The two
extreme conjugacy classes $\eC_\rho(0)$ and $\eC_\rho(\pi/4)$ are $5$-dimensional,
whereas for $0<\vartheta<\pi/4$, $\eC_\rho(\vartheta)$ is $7$-dimensional.
It is not hard to show that $\eC_\rho(\pi/4)$ is a homology sphere, whereas
the integral homology of the other twisted conjugacy classes is given by
\begin{equation*}
  H_p(\eC_\rho(0)) \cong 
  \begin{cases}
    \ZZ & p=0,5~,\\
    \ZZ_2 & p=2~,\\
    0 & p=1,3,4~;
  \end{cases} \qquad\text{and}\qquad
  H_p(\eC_\rho(\vartheta)) \cong 
  \begin{cases}
    \ZZ & p=0,2,5,7~,\\
    0 & p=1,3,4,6~
  \end{cases}
\end{equation*}
for $0<\vartheta<\pi/4$.

The basic tool for computing relative homology groups is the long
exact sequence
\begin{equation*}
  \cdots \to H_p(\eC_\rho) \xrightarrow[\phantom{12}]{i_*} H_p(G) \to 
  H_p(G,\eC_\rho) \to H_{p-1}(\eC_\rho) \xrightarrow[\phantom{12}]{i_*}
  H_{p-1}(G) \to \cdots
\end{equation*}
relating the relative homology of $(G,\eC_\rho)$ to the homologies of
$G$ and $\eC_\rho$.  Since $H_1(G) = H_2(G) = 0$, it follows that
$H_2(G,\eC_\rho) \cong H_1(\eC_\rho)$.  Since $H_1(\eC_\rho)$ vanishes
for all $\eC_\rho=\eC_\rho(\vartheta)$, we conclude that
$H_2(G,\eC_\rho(\vartheta)) = 0$ for all $\vartheta$.

Now let us study $H_3(G,\eC_\rho)$ for each of the twisted conjugacy classes
in turn.

\subsubsection{$\eC_\rho = \eC_\rho(0)$.}

Since $H_3(\eC_\rho) = 0$ and $H_2(G)=0$, the long exact sequence truncates
to a short exact sequence
\begin{equation*}
  0 \to H_3(G) \to H_3(G,\eC_\rho) \to H_2(\eC_\rho) \to 0~.
\end{equation*}
Since $H_3(G) \cong \ZZ$ and $H_2(\eC_\rho)\cong\ZZ_2$, we see that
$H_3(G,\eC_\rho) \cong \ZZ$ and moreover that the map $H_3(G) \to
H_3(G,\eC_\rho)$ is multiplication by $2$.

\subsubsection{$\eC_\rho = \eC_\rho(\pi/4)$.}

Since $H_3(\eC_\rho)=H_2(\eC_\rho)=0$, we have that $H_3(G,\eC_\rho)
\cong H_3(G) \cong \ZZ$.  Therefore $H_3(G,\eC_\rho)$ is generated by
the ``honest'' $3$-cycle generating $H_3(G)$.

\subsubsection{$\eC_\rho = \eC_\rho(\vartheta)$, $0<\vartheta<\pi/4$.}

Since $H_3(\eC_\rho)=0$ and $H_2(G)=0$, we have again the short exact
sequence
\begin{equation*}
  0 \to H_3(G) \to H_3(G,\eC_\rho) \to H_2(\eC_\rho) \to 0~.
\end{equation*}
Since $H_3(G)\cong\ZZ$ and $H_2(\eC_\rho)\cong\ZZ$, we see that
$H_3(G,\eC_\rho) \cong \ZZ \oplus \ZZ$ with generators given by the
generating cycle of $H_3(G)$ and the generating cycle $\Sigma$ of
$H_2(\eC_\rho)$.  Since $H_2(G)=0$, $\Sigma$ bounds in $G$, whence
there exists a $3$-chain $N$ in $G$ such that $\d N = \Sigma$.  The
relative cycle $(N,\Sigma)$ is then one of the generators for
$H_3(G,\eC_\rho)$.  Let us determine the $2$-cycle $\Sigma$ in
$\eC_\rho$.

Consider again the $\SU_2$ subgroup of $\SU_3$ with elements of the
form \eqref{eq:SU2alpha}, and let $\eC_\rho$ be the twisted conjugacy
class of the element $h_\vartheta \in \SU_2$.  Let $\pi: G \to
\eC_\rho$ denote the twisted adjoint action of $G$ on $h_\vartheta$
\begin{equation*}
  \pi(g) = \rho(g) h_\vartheta g^{-1} = \bar g h_\vartheta g^{-1}~.
\end{equation*}
Let $\Sigma\subset \eC_\rho$ denote the submanifold of $\eC_\rho$
obtained by restricting the action to $\SU_2$.  We claim that $\Sigma$
is the generator of $H_2(\eC_\rho)$.  To establish this claim we will
prove that $\int_\Sigma F = 1$, where $F$ is the generator of
$H^2(\eC_\rho)$.  We first need to determine $F$.  We do this as
follows.

The map $\pi:G\to \eC_\rho$ is a fibration with fibre a circle.
Recall (see, for example, \cite{BottTu}) that the Leray spectral
sequence for such a fibration degenerates to a long exact sequence
(the Gysin sequence)
\begin{equation*}
  \cdots \to H^3(\eC_\rho) \xrightarrow[\phantom{12}]{\pi^*} H^3(G)
  \xrightarrow[\phantom{12}]{\pi_*} H^2(\eC_\rho)
  \xrightarrow[\phantom{12}]{e\, \cup} H^4(\eC_\rho) \to \cdots
\end{equation*}
where $\pi^*$ is the pull-back, $\pi_*$ is ``integration along the
fibre'' and $e\, \cup$ is the cup product with the Euler class $e$ of
the circle bundle.  Since $H^3(\eC_\rho)=0=H^4(\eC_\rho)$, we see that
integration along the fibre defines an isomorphism $H^3(G) \cong
H^2(\eC_\rho)$.  This means that the generator of $H^2(\eC_\rho)$ is
given by $F=\pi_*\Omega$, where $\Omega$ is the generator of $H^3(G)$
defined in \eqref{eq:genH3}.  Equivalently, there exists a one-form
$A$ on $G$ such that $\pi^* F = dA$ and such that
\begin{equation}
  \label{eq:AGysin}
\int_{\text{fibre}} A = 1 \qquad\text{and}\qquad  \int_{\SU_2} A
\wedge dA = 1~.
\end{equation}
This means that $\Omega = A \wedge dA + \Omega_h$, where $\Omega_h$ is
horizontal, so that
\begin{equation*}
  \pi_* \Omega = \pi_* (A \wedge dA) =\pi_* (A \wedge \pi^* F) =
  \left( \int_{\text{fibre}} A \right) F = F~.
\end{equation*}

It is easy to find an explicit expression for the one-form $A$
and to check that it satisfies the properties \eqref{eq:AGysin}
above.  Indeed, let $K\subset G$ be the circle subgroup which leaves
invariant the point $h = h_\vartheta$, namely
\begin{equation*}
  K = \left\{ 
      \begin{pmatrix}
        \phantom{-}\cos 2\pi t & \sin 2\pi t & 0 \\
        - \sin 2\pi t & \cos 2\pi t & 0 \\        
        0 & 0 & 1
      \end{pmatrix} \Bigg\vert t \in [0,1] \right\}~.
\end{equation*}
Equivalently, $K = \pi^{-1}(h)$ is the fibre at $h$.  The fibre at the
point $\pi(g) = \bar g h g^{-1}$ is the left translate $gK$ of $K$ by
$g$.  To ensure that the one-form $A$ integrates to $1$ on all the
fibres we choose it to be left-invariant, namely
\begin{equation*}
  A(g) = \Tr X g^{-1}dg
\end{equation*}
for some constant $X \in \su_3$.  A choice of $A$ which satisfies the
properties \eqref{eq:AGysin} is
\begin{equation*}
  A(g) = \frac{1}{4\pi} \Tr \tau_2\,  g^{-1} dg~,
\end{equation*}
where $\tau_2$ is defined in \eqref{eq:taus}.  Indeed, $A$ restricts
to the global angular form $dt$ on the fibres, and on $\SU_2$ to
$-(1/2\pi) \theta^2$, where $\theta^2$ is defined in
\eqref{eq:thetas}.  This implies that $A \wedge dA$ agrees with the
restriction of $\Omega$ to $\SU_2$.  Therefore we could have
restricted the discussion to the fibration $\pi: \SU_2 \to \Sigma$
obtained by restricting the map $\pi$ to $\SU_2$ and we would not
alter any of the conclusions.  In particular, we find that the 2-form
$F = \pi_* \Omega$, where $\pi$ is now restricted to $\SU_2$, is the
generator of $H^2(\Sigma)$, whence $\int_\Sigma F = 1$.

\subsection{Quantisation conditions}

We are now ready to apply the above results to select those
twisted conjugacy classes for which the global worldsheet anomaly
vanishes.  Let us call these classes \emph{admissible}.  Recall
\cite{FSrc} that a submanifold $i:\eC_\rho\into G$ is admissible if
the following three conditions hold:
\begin{enumerate}
\item $H_2(G,\eC_\rho)=0$,
\item the 3-form $H$ in the Wess--Zumino term satisfies $i^* H =
  d\omega$ for some 2-form $\omega$ on $\eC_\rho$; and
\item the relative de~Rham cocycle $\frac{1}{2\pi}(H,\omega)$ is
  integral.
\end{enumerate}
We have proven above that first condition is satisfied; whereas the
second condition follows from an explicit calculation
\cite{AS,Gaw,Q0}.  The third condition means that evaluating the class
$\frac{1}{2\pi}[(H,\omega)] \in H^3(G,\eC_\rho)$ on any relative
3-cycle gives an integer.  As we have seen above, for any twisted
conjugacy class $\eC_\rho$, the homology group $H_3(G,\eC_\rho)$ is
freely generated; hence it is sufficient (and necessary) to check that
\begin{equation}
  \label{eq:quantisation}
  \tfrac{1}{2\pi} \int_N H -   \tfrac{1}{2\pi} \int_{\d N} \omega \in
  \ZZ~,
\end{equation}
where $(N,\d N) \subset (G,\eC_\rho)$ is a generator of
$H_3(G,\eC_\rho)$.  In the previous section we exhibited these
generators in terms of the $\SU_2$ subgroup given by
\eqref{eq:SU2alpha}.  Our strategy will consist in pulling back the
class $\frac{1}{2\pi}[(H,\omega)]$ to $H^3(\SU_2,\Sigma)$ and use the
known results for $\SU_2$.

\subsubsection{$\eC_\rho=\eC_\rho(0)$}  The fundamental 3-cycle $\SU_2$ is twice the 
generator of $H_3(G,\eC_\rho)$; hence evaluating
$\frac{1}{2\pi}[(H,\omega)]$ on $\SU_2$ should give an even integer.
On the other hand, $\frac{1}{2\pi}\int_{\SU_2} H = k$, whence this
class is admissible only when the level $k$ is even.

\subsubsection{$\eC_\rho = \eC_\rho(\pi/4)$.}  The fundamental 3-cycle $\SU_2$ is
the generator, hence $\eC_\rho$ is admissible for all integer levels.

\subsubsection{$\eC_\rho = \eC_\rho(\vartheta)$, $0<\vartheta<\pi/4$.}  There are
two generators of $H_3(G,\eC_\rho)$.  The first generator is the
fundamental cycle $\SU_2$ and the quantisation condition simply says
that the level $k$ is an integer.  For the second generator we can
take any relative cycle $(N,\d N)$ where $\d N = \Sigma$.  Therefore
$\eC_\rho$ is admissible if and only if $\Sigma$ is admissible for
$\SU_2$ at level $k$.  We observe that because complex conjugation is
an inner automorphism for $\SU_2$, $\Sigma$ is a shifted conjugacy
class in $\SU_2$.  Indeed, notice that
\begin{equation*}
  \Sigma = \left\{ \bar g h_\vartheta g^{-1} \vert g \in \SU_2
  \right\} = \left\{ j g j^{-1} h_\vartheta g^{-1} \vert g \in \SU_2
  \right\} = j \eC_\rho(h_{\vartheta+\pi/4})~,
\end{equation*}
where $j$ is defined by
\begin{equation*}
  j := 
  \begin{pmatrix}
    0 & -1 & 0\\ 1 & 0 & 0\\ 0 & 0 & 1
  \end{pmatrix}~,
\end{equation*}
and where $\eC_\rho(h_{\vartheta+\pi/4})$ is the conjugacy class of
$h_{\vartheta+\pi/4}$ in $\SU_2$.  Since $j \in \SU_2$ and $\SU_2$ is
connected, the homotopy invariance of relative (co)homology says that
$j \eC_\rho$ is admissible if and only if so is $\eC_\rho$.  At level
$k$, $\eC_\rho(h_{\vartheta+\pi/4})$ is admissible if and only if
\begin{equation*}
  2\vartheta + \frac{\pi}{2} = \frac{n\pi}{k}
\end{equation*}
for some integer $n$; equivalently,
\begin{equation}
  \label{eq:su2quant}
  \vartheta = \frac{(2n-k)\pi}{2k}\qquad\text{for some integer $n$,}
\end{equation}
which is equivalent to equation \eqref{eq:qmstau}.

\bibliographystyle{utphys}
\bibliography{AdS3,Duality}

\end{document}